\documentclass[useAMS,usenatbib]{mn2e}

\usepackage{graphicx}
\title[Leptonic/hadronic models for MQs]{Leptonic/hadronic models for electromagnetic emission in microquasars: the case of GX 339-4}
\author[G. S. Vila and G. E. Romero]{G. S. Vila$^{1}$\thanks{E-mail:
gvila@iar-conicet.gov.ar} and G. E. Romero$^{1,2}$\\
$^{1}$Instituto Argentino de Radioastronom\'ia, CCT La Plata (CONICET), C.C.5 (1894), Villa Elisa, Buenos Aires, Argentina\\
$^{2}$Facultad de Ciencias Astron\'omicas y Geof\'isicas, Universidad Nacional de La Plata, Paseo del Bosque s/n (1900), La Plata, Argentina}

\begin{document}

\date{}

\pagerange{\pageref{firstpage}--\pageref{lastpage}} \pubyear{2002}

\maketitle

\label{firstpage}

\begin{abstract}

We present a general self-consistent lepto/hadronic jet model for the non-thermal electromagnetic emission of microquasars. The model is applied to the low-mass microquasar (LMMQ) GX 339-4 and predicts its high-energy features. We assume that both leptons and hadrons are accelerated up to relativistic energies by diffusive shock acceleration, and calculate their contribution to the electromagnetic spectrum through all main radiative processes. The radiative contribution of secondary particles (pions, muons and electron-positron pairs) is included. We use a set of simultaneous observations in radio and X-rays to constrain the model parameters and find the best fit to the data. We obtain different spectral energy distributions that can explain the observations, and make predictions for the high-energy emission. Observations with gamma-ray instruments like Fermi can be used to test the model and determine the proton content of the jets. Finally, we estimate the positron injection in the surrounding medium. Our findings support the suggested association between LMMQs and the observed distribution of the 511 keV line flux observed by INTEGRAL.    

\end{abstract}


\section{Introduction}

The low-mass microquasar GX 339-4 was discovered in 1972 by the satellite OSO--7 (Markert, Canizares \& Clark 1973). Since then, it has been extensively observed at all wavelengths from radio to X-rays, and detected in all the canonical spectral states of X-ray binaries. Little is known, however, about the remaining characteristics of the binary system.

Based on modulations in the optical photometry, Callanan et al. (1992) inferred an orbital period of 14.8 hs, later confirmed by Buxton \& Vennes (2003) using optical spectroscopy. Further optical spectroscopic measurements and analysis of long-term X-ray light curves showed no evidence of this modulation, revealing instead a periodicity of $\sim$1.75 days (Hynes et al. 2003; Levine \& Corbet 2006). The first estimates of the distance to GX 339-4 placed the system at $d\sim1.3-4$ kpc, see Zdziarski et al. (1998) and references therein. This result was later revised by Zdziarski et al. (2004), who concluded that the minimum distance lay in the range $6.7$ kpc $\la d_{\rm{min}}\la 9.4$ kpc. They favoured a location in the galactic bulge at $\sim8$ kpc. However, a study of absorption lines performed by Hynes et al. (2004) suggests that a location in the far side of the Galaxy at a distance $d>15$ kpc cannot be completely ruled out.
  
The emission in the optical band is dominated by the accretion flow (Imamura et al. 1990), preventing direct observation of the secondary star even when the system is going through the very low X-ray luminosity state. The first detection of the star was made by Hynes et al. (2003) during the X-ray outburst observed in 2002. The mass and spectral type of the donor star have not been firmly established yet. According to Hynes et al. (2004), an orbital period of $\sim$1.7 days implies a low density for the companion of $\sim0.06$ g cm$^{-3}$. This probably corresponds to a low-mass subgiant of spectral type G or F, depending on the assumed distance.  Following this idea, Mu\~noz-Darias, Casares \& Mart\'inez-Pais (2008) suggested that the star is a ``stripped-giant'', in which mass loss is due to the burning of an hydrogen shell.  In this model, the mass of the secondary must be in the range $0.166M_\odot<M_2<1.1M_\odot$. From this result, the authors constrained the mass of the compact object to be $M_{\rm{BH}}>6M_\odot$ or even $M_{\rm{BH}}>8.6M_\odot$, for a mass of the secondary near the lower or upper limit, respectively. These values strongly support the idea that the compact object is a black hole (see also Hynes et al. 2003).

GX 339-4 has been observed in radio, infrared, optical and X-ray wavelengths, sometimes simultaneously or quasi--simultaneously (Hannikainen et al. 1998; Wilms et al. 1999; Nowak, Wilms \& Dove 2002; Homan et al. 2005). The source goes through all the spectral states of X-ray binaries: low-hard, high-soft, very high state, intermediate state and quiescence (McClintock \& Remillard 2006). It frequently displays outbursts associated with state transitions, episodes during which the X-ray luminosity can reach peaks of $L_X = 10^{37-38}$ erg s$^{-1}$ for an assumed distance of 6 kpc (Homan et al. 2005; Yu et al. 2007). It was after the X-ray outburst of 2002 that Gallo et al. (2004) imaged for the first time a  relativistic radio jet on $\sim10^3$ AU scales in the system (see also Corbel et al. 2000). The detection of the jet confirmed that GX 339-4 is a microquasar. 

The hard X-ray emission during the low-hard state of microquasars and black hole binaries is generally thought to have its origin in a hot corona that surrounds the black hole. The observed power-law spectrum is explained through Compton up-scattering of accretion disc photons by hot electrons in the corona. Some of these up-scattered photons in turn excite iron nuclei in the disc material, giving rise to the appearance of a K$\alpha$ line at $\sim6.5$ keV, superimposed on the power-law continuum; for evidence supporting the presence of the Fe K$\alpha$ line in GX 339-4, see for example Dunn et al. (2008). However, Corbel et al. (2003) showed that in GX 339-4 the radio and X-ray fluxes are tightly correlated, $F_{\rm{radio}}\propto F_X^{0.7}$. This suggests that the emission in both bands might have a common origin in synchrotron radiation produced by non-thermal electrons in the jet, and not in the corona (Corbel \& Fender 2002; Corbel et al. 2003). This idea was explored by Markoff et al. (2003) and Markoff, Nowak \& Wilms (2005), who applied a jet model to fit the observations. They showed that synchrotron radiation of relativistic electrons in the base of a jet can explain both the radio and X-ray spectra and reproduce the observed correlation. This model, however, is  purely leptonic: the contribution to the radiation output of relativistic protons in the jet is not taken into account. 

In this work we present a lepto-hadronic model for the broadband electromagnetic spectrum of GX 339-4. In our model, protons and electrons are accelerated up to relativistic energies. The hadronic radiative contribution extends well into the gamma-ray band, since typical maximum proton energies are much higher than those of electrons. We expect that our predictions can be tested in the near future with the data collected by instruments like the Fermi gamma-ray satellite and by atmospheric Cherenkov arrays like the High Energy Stereoscopic System II (HESS II). 

Along with electromagnetic radiation, the creation of electron-positron pairs is a necessary result of relativistic particle interactions. They are injected, for example, through photon-photon annihilation and as a by-product of hadronic interactions. 

Recently, measurements carried out with the INTEGRAL Spectrometer (SPI) instrument of the INTEGRAL satellite have allowed to complete a detailed map of an extended region of emission line at 511 keV in the Galaxy (Weidenspointner et al. 2008). These observations confirm the diffuse (rather than point-like) distribution of the line, with bright emission around the galactic center (flux $\sim 10^{-3}$ cm$^{-2}$ s$^{-1}$), and a clear, asymmetric (toward negative longitude values) disc component. The disc total flux is $\sim 1/5$ of the bulge flux. The disc emission is detected up to scales of $\sim 20^{\circ}$ from the galactic center. 

Different types of positron sources have been suggested in the literature, including pulsars, the massive black hole at the galactic center, microquasars, nucleosynthesis events, and extended processes like cosmic ray nuclear reactions and dark matter decay. The fact that both bulge and disc emission are clear, seems to disfavour the latter two possibilities as the main positron sources. Microquasars, in particular those with a low-mass donor star, seem to be a particularly appealing possibility, given the spatial correlation of the line emission with the overall distribution of low-mass X-ray binaries in the Galaxy (Weidenspointner et al. 2008) and energetic arguments (Guessoum, Jean, \& Prantzos 2006). 

Up to the present, none of the microquasar models available presents quantitative predictions of the positron production rate, although in Romero \& Vila (2008, 2009) the radiative output of positrons was taken into account. The recent findings described above makes it timely an exploration of the possibilities for electron-positron production in self-consistent models for microquasar jets. In the present paper we have devoted a section to estimate the positron production rate in systems like GX 339-4.

The article is organized as follows: in Section \ref{Model} we present the basis of the model, describing in detail the hypothesis made about the geometry and energetics of the system. We also describe the calculation of the relativistic particle distributions (for primary and secondary particles) and the radiative processes. In Section \ref{General results}, we discuss the observational data sets and the constraints they impose on the model parameters. We also present the best-fitting spectral energy distributions obtained and analyse the results. Our estimations for the positron production rate are described in Section \ref{positrons}. Finally, in Section \ref{Conclusions}, we summarise our results and we compare them with those of previous works. We close discussing the predictions for the very high-energy emission. 

\section{Model}
\label{Model}

\subsection{Basic picture}

\begin{figure}
\centering
\includegraphics[width=0.4\columnwidth, keepaspectratio]{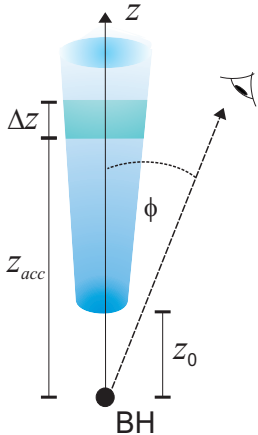}
\caption{Scheme of the jet. It is launched at a distance $z_0$ of the black hole (BH), and acceleration of particles takes place further away at $z_{\rm{acc}}$. The width of the acceleration region is indicated as $\Delta z$. The jet axis makes an angle $\phi$ with the line of sight.}
\label{jet-sketch}
\end{figure}

The jet model applied here is based on the model developed in Romero \& Vila (2008). A sketch of the system is shown in Figure \ref{jet-sketch}. The jet is supposed to be conical, and launched at a distance $z_0=50R_{\rm{g}}$ from the compact object, where $R_{\rm{g}}=GM_{\rm{BH}}/c^2$ is the gravitational radius of the black hole. The initial radius of the jet is taken to be a fraction $\chi$ of the value of the injection distance, $r_0=\chi z_0$. Following Falcke \& Biermann (1995), we relate the jet power to the Eddington luminosity of the black hole as

\begin{equation}
	L_{\rm{jet}}=\frac{1}{2}q_{\rm{jet}}L_{\rm{Edd}},
	\label{Ljet-Laccr} 
\end{equation}

\noindent where $L_{\rm{Edd}}\approx 1.3\times 10^{38}(M_{\rm{BH}}/M_\odot)$ erg s$^{-1}$. The factor $1/2$ accounts for the existence of a counterjet of equal power. \\

At an arbitrary distance $z$ from the compact object, the total energy budget of the jet can be roughly divided into magnetic energy, bulk kinetic energy and particle internal energy\footnote{Strictly, $L_{\rm{jet}}>L_{\rm{jet}}(z)$ since part of $L_{\rm{jet}}$ is dissipated as radiation.}:

\begin{equation}
	L_{\rm{jet}}(z)\approx L_{\rm{B}}(z) + L_{\rm{kin}}^{\rm{bulk}}(z) + L_{\rm{int}}(z). 
	\label{jet_energy_budget1} 
\end{equation}

\noindent If the plasma is ejected by some kind of magnetocentrifugal mechanism, we can assume that at the base of the jet the entire jet power is in the form of magnetic energy, $L_{\rm{jet}}\approx L_{\rm{B}}(z_0)$. This allows to estimate the value of the magnetic field $B_0=B(z_0)$ by equipartition between the magnetic energy density and the total energy density,

\begin{equation}
	\frac{B_0^2}{8\pi}=\frac{L_{\rm{jet}}}{\pi r_0^2v_{\rm{jet}}},
	\label{equipartition} 
\end{equation}

\noindent where $v_{\rm{jet}}$ is the bulk velocity of the outflow. After the launching point the magnetic field decreases with the distance $z$ to the compact object as

\begin{equation}
	B(z)=B_0 \left(\frac{z_0}{z}\right)^m,
	\label{B(z)} 
\end{equation}   

\noindent with $1\leq m \leq 2$ (Krolik 1999). Magnetic energy is then converted into bulk kinetic energy (and a fraction of this is dissipated at shocks and converted into internal energy of relativistic particles, see below), in such a way that equipartition holds only at the base of the jet. 

A few remarks are in order here. In the region $z<z_{\rm{acc}}$ the outflow is a Poynting-dominated flux, formed by a thermal plasma that accelerates as it gets energy from the ordered magnetic field. Thermal electrons and ions in the ``pre-acceleration'' region move approximately with the jet bulk velocity and have an energy $\sim\Gamma_{\rm{jet}}mc^2$. The plasma is neutral, but almost all the bulk kinetic energy of the jet is carried by the ions. No significant electron synchrotron radiation is expected from this region. At $z\approx z_{\rm{acc}}$ shocks perturb the field, and a fraction of the energy associated with the bulk motion of the plasma is converted into internal energy of relativistic non-thermal particles. In this region synchrotron cooling becomes efficient, especially for electrons. This situation is different from the case of high-mass microquasars, where ``cold'' relativistic electrons can cool by inverse Compton interactions with the stellar photon field. A similar scenario arises in binary systems formed by a pulsar and a luminous Be star, such as PSR B1259-63/SS2883, see Khangulyan et al. (2007). 

Most of the bulk kinetic energy of the jet is carried by a thermal plasma. If shock waves propagate through some region of the outflow, the suprathermal tail of the Maxwellian particle distribution can be accelerated up to relativistic energies by diffusion across the shock front (see, for instance, Drury 1983 and references therein). The physical conditions for an efficient acceleration are not clear.  For the plasma to be mechanically compressible and allow the formation of shocks, the magnetic energy density $U_{\rm{B}}=B^2/8\pi$ must be in sub-equipartition with the bulk kinetic energy density $U_{\rm{k}}$ of the plasma (see Komissarov et al. 2007 for a discussion on this topic). Therefore, the acceleration region has to be located at a distance $z_{\rm{acc}}$ from the black hole such that

\begin{equation}
	U_{\rm{B}}(z_{\rm{acc}})< \,U_{\rm{k}}(z_{\rm{acc}}).
	\label{sub-equipartition} 
\end{equation}  

\noindent The kinetic energy density of the jet can be written as  

\begin{equation}
U_{\rm{k}}= n(z)E_p^{\rm{kin}},
	\label{matter_energy_density} 
\end{equation}  

\noindent where  $n(z)$ is the cold particle density, and $E_p^{\rm{kin}}$ is the relativistic kinetic energy of a proton that moves with the jet bulk velocity,

\begin{equation}
E_p^{\rm{kin}} = (\Gamma_{\rm{jet}}-1)m_pc^2.
	\label{thermal_energy} 
\end{equation}  

The presence of shocks may not suffice to accelerate particles efficiently. According to Gaisser (1990), for diffusive shock acceleration to work, the ram pressure in the acceleration region must dominate  over the magnetic pressure. This condition can be written as

\begin{equation}
	U_{\rm{B}}(z_{\rm{acc}})< \,\frac{2}{3}U_{\rm{m}}(z_{\rm{acc}}),
	\label{sub-equipartition_2} 
\end{equation}

\noindent where $U_{\rm{m}}$ is the internal matter  energy density. For a cold-proton-dominated jet,  $U_{\rm{m}}$ can be calculated as in Bosch-Ramon, Romero \& Paredes (2006),

\begin{equation}
U_{\rm{m}}=\frac{\dot{M}_{\rm{jet}}}{\pi r_{\rm{jet}}^2v_{\rm{jet}}m_p}\bar{E}_p^{\rm{kin}}.
	\label{thermal_energy_density} 
\end{equation} 

\noindent Here $\dot{M}_{\rm{jet}}$ is the jet mass flow,

\begin{equation}
\dot{M}_{\rm{jet}}\approx \frac{L_{\rm{jet}}}{\Gamma_{\rm{jet}}c^2},
	\label{jet_mass_accretion_rate} 
\end{equation} 

\noindent and $\bar{E}_p^{\rm{kin}}$ is the classical kinetic energy of a thermal proton,

\begin{equation}
\bar{E}_p^{\rm{kin}}=\frac{1}{2}m_pv_p^2.
	\label{classical_kinetic_energy} 
\end{equation} 

The mean velocity of the particles was taken to be equal to the lateral expansion velocity of the jet, $v_p=v_{\rm{exp}}=\chi v_{\rm{jet}}$, that is of the order of the speed of sound in the plasma. Condition (\ref{sub-equipartition_2}) is stronger than (\ref{sub-equipartition}), in the sense that if the former is fulfilled, so is the latter. In either case, the location of the innermost acceleration region can be determined demanding that the appropriate condition is satisfied,

\begin{equation}
U_{\rm{B}}=\rho\, U_{\rm{(k,m)}},
	\label{condition_z_acc} 
\end{equation} 

\noindent with $\rho<1$.

We assume that the jet is dynamically dominated by cold matter. The total power $L_{\rm{rel}}$ injected in relativistic particles is only a small fraction of the jet power,

\begin{equation}
	L_{\rm{rel}}= q_{\rm{rel}}L_{\rm{jet}}.
	\label{Lrel-Ljet} 
\end{equation}

\noindent This energy is shared between relativistic protons and leptons, $L_{\rm{rel}}=L_p+L_e$. We relate the energy budget of both species as

\begin{equation}
	L_p = a\,L_e.
	\label{Lp-Le} 
\end{equation}

\noindent The parameter $a$ remains free in our model but, as we are interested in jets with a relevant hadronic content, we kept $a\ge1$ throughout. 
 
\subsection{Fundamental equations}

\subsubsection{Particle distributions}

The differential spectrum of particles accelerated by diffusion through shock waves is a power-law in energy, and the particle injection function can be written as

\begin{equation}
Q\left(E\right) = Q_0\,E^{-\alpha}.
	\label{injection-function} 
\end{equation}  

\noindent The normalization constant $Q_0$ is calculated from the total injected power,

\begin{equation}
 L_{(e,p)}=\int_{V}\mathrm{d}^3r\int_{E^{\rm{min}}}^{E^{\rm{max}}}\,E\,\,Q_{(e,p)}(E)\,\,\mathrm{d}E,
\label{normalization_constant}
\end{equation}  

\noindent where $V$ is the volume of the acceleration region. Particles can gain energy up to a certain value $E^{\rm{max}}$ for which the sum of the cooling rates 

\begin{equation}
	t^{-1}_{\mathrm{cool}, i} = -\frac{1}{E}\left.\frac{dE}{dt}\right|_{i}
	\label{energy_loss_rate}
\end{equation} 

\noindent over all the relevant processes of energy loss $i$, equals the acceleration rate (Aharonian 2004),

\begin{equation}
	t^{-1}_{\rm{acc}}=\eta\frac{ecB}{E}.
	\label{acc-rate}
\end{equation}

\noindent The parameter $\eta<1$ characterises the efficiency of the acceleration mechanism. 

The steady-state energy distributions of relativistic particles $N\left(E\right)$ is then calculated solving the transport equation in the one--zone approximation in the jet co--moving reference frame (where the distributions are supposed to be isotropic), 

\begin{equation}
	\frac{\partial}{\partial E}\biggl[\left.\frac{dE}{dt}\right|_{\rm{total}}N(E)\biggr]+\frac{N(E)}{T}=Q(E).
	\label{transp_eq}
\end{equation} 

\vspace{0.2cm}

\noindent The time $T$ is the characteristic time of ``catastrophic'' non-radiative particle losses: processes through which particles are removed from the system, for example by escape or decay.  As protons and electrons are stable particles, $T$ is simply equal to the escape time-scale from the acceleration region,

\begin{equation}
T_{(p,e)} = t_{\rm{esc}}\approx \left(\frac{\Delta z}{v_{\rm{jet}}}\right).
	\label{escape-time}
\end{equation}

\vspace{0.2cm}

Equation (\ref{transp_eq}) neglects effects of convection and diffusion, and therefore it is only valid in a thin region of the jet. Its exact analytical solution can be found, for example, in Khangulyan et al. (2007).

\subsubsection{Radiative processes}

We consider several mechanisms of interaction of the relativistic particles with the magnetic field, photons and matter in the jet. Both protons and leptons interact with the magnetic field emitting synchrotron radiation. Electrons also radiate by relativistic Bremsstrahlung as they are accelerated in the mean electrostatic field of the ions in the jet plasma, and by inverse Compton scattering (IC) against the synchrotron photon field (Synchrotron Self Compton process, SSC). Interaction of protons with the synchrotron field ($p\gamma$) proceeds through photopair production 

\begin{equation}
p+\gamma\rightarrow p+e^-+e^+,
	\label{photopair}
\end{equation} 

\noindent and photomeson production

\begin{equation}
p+\gamma\rightarrow p+a\pi^0+b\left(\pi^++\pi^-\right),
	\label{photomeson1}
\end{equation}  

\begin{equation}
p+\gamma\rightarrow n+\pi^++a\pi^0+b\left(\pi^++\pi^-\right).
	\label{photomeson2}
\end{equation}   

\noindent An approximation for the cross section for photopair production is given in Maximon (1968). If $\epsilon^\prime=x'm_e c^2$ is the photon energy in the proton rest frame, then for $x'\la2$

\begin{equation}
\sigma_e = 1.2\times\left(\frac{x'-2}{x'}\right)^3\left[1+\frac{1}{2}\left(\frac{x'-2}{x'+2}\right)+...\right]\rm{mb},
	\label{sigma_photopair1}
\end{equation}   

\noindent whereas for $x'\ga2$

\begin{equation}
\sigma_e = 0.58\times\left[3.1\ln 2x'-8.07+...\right]\rm{mb}.
	\label{sigma_photopair2}
\end{equation}

\noindent In the case of photomeson interactions, the cross section $\sigma_\pi$ can be roughly approximated as a step function (Atoyan \& Dermer 2003),

\begin{displaymath}
\sigma_{\pi}(\epsilon^\prime)\approx\left\{\begin{array}{ll}
340\,\mu\mathrm{barn} \quad 200\,\mathrm{MeV}\leq\epsilon^\prime\leq500\,\mathrm{MeV} \\[0.3cm]
120\,\mu\mathrm{barn} \quad \epsilon^\prime>500\,\mathrm{MeV},
\end{array} \right.
\label{sigma_phtomeson}
\end{displaymath}

\noindent Relativistic protons also produce pions by interaction with the jet matter field through proton-proton $(pp)$ collisions,

\begin{equation}
p+p\rightarrow p+p+a\pi^0+b\left(\pi^++\pi^-\right),
	\label{pp1}
\end{equation}  

\begin{equation}
p+p\rightarrow p+n+\pi^++a\pi^0+b\left(\pi^++\pi^-\right),
	\label{pp2}
\end{equation}    

\noindent and

\begin{equation}
p+p\rightarrow n+n+2\pi^++a\pi^0+b\left(\pi^++\pi^-\right).
	\label{pp3}
\end{equation}  

\noindent For high proton energies, the probability of creation of the three pion species is almost the same. The total inelasticity of the process is $\sim0.5$; most of the kinetic energy lost by the proton is carried away by only one or two leading pions. The inelastic $pp$ cross section can be accurately approximated as (Kelner, Aharonian \& Bugayov 2006)

\begin{eqnarray}
	\sigma_{\rm{inel}}\left(E_p\right)=\left(34.3+1.88\ln\frac{E_p}{\rm{1TeV}}+0.25\ln^2\frac{E_p}{\rm{1TeV}}\right)\,\times \;\nonumber\\  \left[1-\left(\frac{E_{\rm{th}}}{E_p}\right)^4\right]^2\,\rm{mb},
	\label{pp3b}
\end{eqnarray}

\vspace{0.2cm}

\noindent where  $E_{\rm{th}}=1.22\,$ GeV is the proton threshold energy for $\pi^0$ production. The pion multiplicities $a$ and $b$ have power-law dependences on the relativistic proton energy, $a,b\propto E_p^{-\kappa}$ with $\kappa\sim1/4$ (Mannheim \& Schlickeiser 1994).\\

\noindent The gamma-ray output from proton-proton and photomeson interactions is due to the decay of neutral pions, 

\begin{equation}
\pi^0\rightarrow 2\gamma.
	\label{pp4}
\end{equation}

Besides radiative cooling, particles also lose energy through adiabatic losses, since they exert work on the jet boundary surface. The expression for the adiabatic cooling rate and for all the radiative mechanisms of energy loss are compiled in Romero \& Vila (2008) and Reynoso \& Romero (2009).

We calculate the contribution of each radiative process to the luminosity. For synchrotron radiation, relativistic Bremsstrahlung, and inverse Compton scattering (in both Thomson and Klein-Nishina regimes) we use the formulae given in Blumenthal \& Gould (1970). The gamma-ray luminosities from proton-proton and photomeson interactions are calculated following Kelner, Aharonian \& Bugayov (2006) and Kelner \& Aharonian (2008), respectively. 

All the luminosities, except those from process of interactions with matter, are better calculated in the jet co--moving reference frame. The corresponding luminosity in the observer frame is then obtained applying the Doppler boosting factor $D(\phi)$,

\begin{equation}
D=\left[\Gamma_{\rm{jet}}\left(1-\beta_{\rm{jet}}\cos\phi\right)\right]^{-1},
	\label{boost-factor}
\end{equation}

\noindent where $\phi$ is the viewing angle (the angle between the jet axis and the line of sight) and  $\beta_{\rm{jet}}=v_{\rm{jet}}/c$. Denoting with primes the magnitudes in the co--moving frame, for an approaching jet the luminosity transforms as  

\begin{equation}
L_\gamma(E_\gamma)=D^2\,L_\gamma^\prime(E_\gamma^\prime),
	\label{boost-luminosity}
\end{equation}

\noindent whereas the photon energy in the observer frame is

\begin{equation}
E_\gamma=DE_\gamma^\prime.
	\label{boost-energy}
\end{equation}

\subsubsection{Secondary particles}

Charged pions, muons and electron-positron pairs produced in proton-proton and proton-photon interactions can also contribute to the electromagnetic emission spectrum. If the magnetic field and/or the photon and matter fields in the jet are strong enough, $\pi^\pm$ and $\mu^\pm$ can significantly cool through synchrotron radiation, inverse Compton scattering and even pion-proton collisions $(\pi p)$ before decaying. Charged pions decay creating muons and neutrinos,

\begin{equation}
\pi^-\rightarrow \mu^-+ \bar{\nu}_\mu,
	\label{pi-decay1}
\end{equation}

\begin{equation}
\pi^+\rightarrow \mu^++ \nu_\mu.
	\label{pi-decay2}
\end{equation}

Muons in turn decay yielding electrons, positrons and more neutrinos,

\begin{equation}
\mu^-\rightarrow e^-+\nu_\mu+\bar{\nu}_e,
	\label{mu-decay1}
\end{equation}

\begin{equation}
\mu^+\rightarrow e^++\bar{\nu}_\mu+\nu_e.
	\label{mu-decay2}
\end{equation}

The pion injection function $Q_{\pi^\pm}$ for pions created in $pp$ collisions is given in Kelner, Aharonian \& Bugayov (2006); for pions injected through photomeson interactions we apply the $\delta$-functional approximation, see Atoyan \& Dermer (2003). From  $Q_{\pi^\pm}$, we calculate the muon injection function $Q_{\mu^\pm}$ using the formalism presented in Lipari, Lusignoli \& Meloni (2007).

Electron-positron pairs created by muon decay and those directly injected through photopair interactions are also a source of electromagnetic radiation. We calculate the pair injection function from muon decays following Schlickeiser (2002). To estimate the pair output $Q_{e^\pm}$ from photopair production, we apply the formulae given in Chodorowski, Zdziarski \& Sikora (1992) and M\"ucke et al. (2000). Finally, there is another source of pair injection due to photon-photon ($\gamma\gamma$) annihilation,

\begin{equation}
\gamma+\gamma\rightarrow e^++e^-.
	\label{gamma-gamma}
\end{equation}

\noindent The corresponding pair source function is given in B\"ottcher \& Schlickeiser (1997). The process of $\gamma\gamma$ annihilation is also a photon sink, and can eventually modify the production spectrum. Its effect on the escape of photons is discussed in the next section.

The steady state distribution of secondary pions, muons and electron-positron pairs is calculated solving the transport equation (\ref{transp_eq}), in the same way  as for the primary particles. In the case of $\pi^\pm$ and $\mu^\pm$, the decay time must be included in $T$ along with the escape time,

\begin{equation}
T^{-1}_{(\pi,\mu)} = t^{-1}_{\rm{esc}}+t^{-1}_{\rm{dec}}. 
	\label{decay-escape-time}
\end{equation}

In the jet co--moving reference frame, the decay time $t_{\rm{dec}}$ is given by

\begin{equation}
t_{\rm{dec}}= \tau_{(\pi,\mu)}\,\gamma_{(\pi,\mu)}.
	\label{decay-time}
\end{equation}

\noindent Here $\gamma_{(\pi,\mu)}=E_{(\pi,\mu)}/m_{(\pi,\mu)}c^2$ is Lorentz factor of the particles and $\tau_{(\pi,\mu)}$ is the decay time in their rest frame, $\tau_{\pi}=2.6\times10^{-8}$ s and $\tau_{\mu}=2.2\times10^{-6}$ s. 

\subsubsection{Attenuation of the production spectrum}

As remarked above, gamma-ray photons can annihilate against low-energy radiation to create electron-positron pairs. For a gamma ray of energy $E_\gamma$ to interact with a photon of energy $\epsilon$, their energies must exceed the threshold 

\begin{equation}
\epsilon E_\gamma\left(1-\cos\theta\right)=2m_e^2c^4. 
	\label{gamma-gamma-threshold}
\end{equation}

\noindent The angle $\theta$ is that between the momenta of the colliding photons. According to equation (\ref{gamma-gamma-threshold}), TeV gamma rays can be strongly absorbed by infrared photons produced, for example, by electron synchrotron radiation in the jet. Thus, this process of internal absorption can modify the shape of the high-energy region of the production spectrum. The probability of absorption of a gamma ray can be quantified through the opacity  $\tau_{\gamma\gamma}$, defined as

\begin{eqnarray}
	\tau_{\gamma\gamma}(E_{\gamma})=\frac{1}{2}\int_{l}\,\,\int^{\epsilon_{\rm max}}_{\epsilon_{\rm th}}\int^{u_{\rm max}}_{-1} \,\sigma_{\gamma\gamma}(E_{\gamma},\epsilon,u)\,n(\epsilon)\,\times  \;\nonumber\\ \,\,\,\,\,(1-u)\,\mathrm{d}u \; \mathrm{d}\epsilon \;\mathrm{d}l.
	\label{tau_gamma-gamma}
\end{eqnarray}

\noindent Here $\sigma_{\gamma\gamma}$ is the angle--averaged cross section (Gould \& Schr\'eder 1967), $n(\epsilon)$ is the distribution of target photons, $u=\cos\theta$, and $l$ is the length of the path traversed by the gamma ray from the emission site to the observer. The shape of the modified luminosity $\widetilde{L}_\gamma $ can be estimated from the primary production spectrum as

\begin{equation}
\widetilde{L}_\gamma\left(E_\gamma\right)=\exp\left(-\tau_{\gamma\gamma}\right)\,L_\gamma\left(E_\gamma\right). 
	\label{attenuated-luminosity}
\end{equation}

\noindent This provides a zeroth-order approximation to the final emission spectrum, mainly indicating those energy ranges where the emission is suppressed. A more refined treatment of the problem requires solving a set of coupled equations to calculate self-consistently the equilibrium distribution of particles and photons, or to solve the equations of an electromagnetic cascade, if it develops. However, the strong magnetic field suppresses IC cascades through fast cooling by the synchrotron channel (see Khangulyan, Aharonian \& Bosch-Ramon 2008). Cascades are expected to play a more important role in high-mass X-ray binaries, see for example Orellana et al. (2007).

\section{General results}
\label{General results}

\subsection{Application to GX 339-4}

GX 339-4 was extensively observed simultaneously in radio and X-rays during the low-hard state (LHS) in 1997, 1999 and 2002. For some of these observations, simultaneous near infrared (NIR) and optical data are also available. The 1997 and 1999 radio observations were carried out with the Australia Telescope Compact Array (ATCA) and the Molonglo Observatory Synthesis Telescope (MOST), and are described in detail in Corbel et al. (2000). The radio, NIR and optical data from 2002 is presented in Homan et al. (2005). The X-ray data were collected with the Rossi X-ray Timing Explorer (RXTE), and are compiled in Wilms et al. (1999), Nowak, Wilms \& Dove (2002), Corbel et al. (2003) and Homan et al. (2005). We refer the reader to these works for model assumptions and other details of the data extraction in each case. No further data reductions were performed for this paper.\footnote{Calibration and data reduction algorithms have been updated since the data presented here were reduced. Reprocessing the data might result in changes in the slope of the X-ray spectrum. However, we do not intend to perform detailed fits to the spectrum but to show that the observations can be accounted for by a lepto-hadronic jet model, as an alternative to purely leptonic models. Using the same data allows to compare the results of our model with those of previous works.} Additional information on the particular data sets used here is presented in Table \ref{obs_data}. 

\begin{table*}
 \centering
 \begin{minipage}{140mm}
  \caption{Observational data used in the model fits.}
  \label{obs_data} 
  \begin{tabular}{@{}cccccccc@{}}
  \hline
  Observation  &  Date       & \multicolumn{3}{c}{X-ray flux }                         & \multicolumn{3}{c}{Radio flux }  \\
               & (y/m/d)     & \multicolumn{3}{c} {($10^{-9}$ erg cm$^{-2}$ s$^{-1}$)} & \multicolumn{3}{c}{(mJy)}        \\   
 \hline                       
               &                      &  3-9 keV & 9-20 keV & 20-200 keV              & 0.8 GHz & 4.8 GHz  & 8.6 GHz      \\
 \hline
  Obs1         &  1997/02/03           &  1.06    & 1.02     & 4.95                   & 7.0      & -      & 9.1        \\
  Obs2         &  1999/04/02           &  0.49    & 0.48     & 2.75                   & -        & 4.8    & 5.1        \\  
  Obs3         &  1999/06/25           &  0.059   & 0.052    & $<$0.29                & -        & 0.14   & 0.34       \\
  Obs4         &  $\quad$1999/08/28$^{(a)}$   &  0.037   & $<$0.01  & $<$0.17         & -        & -      & 0.35        \\  
  \hline	
               &             & {X-ray flux }                        & \multicolumn{3}{c}{IR/optical magnitudes }  & Radio flux \\													                      &             & {($10^{-9}$ erg cm$^{-2}$ s$^{-1}$)} &  \multicolumn{3}{c}{ }                      &  (mJy)     \\  
   \hline             			               
  						 &             & {$3-300$ keV}                       & $m_H$    & $m_I$    & $m_V$                  &   4.8 GHz   \\                          
  \hline
  Obs5         &  2002/03/22 &    15.1                             &   11.7   &  14.1    &  15.6                  &    13.3
  \\     
 \hline
\end{tabular}
\\ $^{(a)}$ Radio data from 1999/09/01.
\\
X-ray data were collected with the Proportional Counter Array (PCA) and High Energy X-ray Timing Experiment (HXTE) instruments of the RXTE satellite. The HXTE measurements are normalized to PCA flux levels. Optical and IR photometry was obtained with the Yale-AURA-Lisbon-Ohio State (YALO) telescope. The radio flux density at 0.8 GHz was obtained with MOST, and at 8.6 GHz and 4.8 GHz with ATCA. From Corbel et al.(2000), Nowak, Wilms \& Dove (2002), Corbel et al. (2003) and Homan et al. (2005).
\end{minipage}
\end{table*}

We placed the injection point of the jet at a distance $z_0=50R_{\rm{g}}\approx4.5\times10^7$ cm from the black hole, and fixed the ratio $\chi=r_0/z_0=0.1$ (Romero \& Vila 2008). This gives a jet half-opening angle of $\approx6^\circ$. The position of the acceleration region was determined from equation (\ref{sub-equipartition_2}), demanding sub-equipartition of the magnetic energy density with respect to the jet internal matter energy density at $z=z_{\rm{acc}}$. A second scenario in which $z_{\rm{acc}}$ is fixed from equation (\ref{sub-equipartition}) was also considered. We constrained the ratio of energy densities to be $\rho<1$, but otherwise it was left as a free parameter. In any case the acceleration/emission region was taken to be thin, $\Delta z=4z_{\rm{acc}}$, as required for the one-zone approximation to be valid. We adopted several decay prescriptions for $B(z)$, with $m=$1.2, 1.5, 1.8 and 2.

Obs1, Obs2 and Obs5 correspond to the end of the low-hard state, when the source was highly luminous. Assuming a conservative value of $d=6$ kpc for the distance, the observed X-ray fluxes yield luminosities of up to $L_X\approx10^{37}$ erg s$^{-1}$. This places some constraints on the value of the parameters that determine the energetics in our model. Only a small fraction of the jet power is carried by relativistic particles, otherwise the outflow could not be confined; we fixed $q_{\rm{rel}}=0.1$ in equation (\ref{Lrel-Ljet}). In a  model with equipartition between hadrons and leptons ($a=1$), half of this energy is given to relativistic electrons. If the observed X-ray flux is due to electron synchrotron radiation, this implies at least a total jet power $L_{\rm{jet}}\approx2\times10^{38}$ erg s$^{-1}$. This is a significant fraction of the Eddington luminosity of a black hole of $M_{BH}=6M_\odot$, $L_{\rm{Edd}}\approx 7.8\times 10^{38}$ erg s$^{-1}$. If part of the accretion power is radiated outside the jet and part advected onto the black hole, the accretion rate required to account for the observations must be very near the Eddington limit. 

An accretion model that could apply to powerful sources or high luminosity states has been proposed by Bogovalov \& Kelner (2005). They showed that, along with the standard thin disc solution of Shakura \& Sunyaev (1973), there exists another accretion regime in which the disc is radiatively very inefficient, even for high accretion rates. In this solution, known as the ``dissipationless disc model'', a magnetized plasma falls onto a central object. The plasma is attached to the magnetic field, and angular momentum is removed from the system not by viscosity effects, but it is carried away by matter itself (see also the Advection-Dominated Inflow-Outflow Solution model of Blandford \& Begelman 1999, for early ideas regarding this accretion regime). In fact, the model predicts that the mass advection rate vanishes at $r=0$, and all the infalling matter is ejected. In this way, most of the accretion power could be directly channeled into the jets. This model could account for the observations of very powerful jets and low-luminosity discs in extreme systems such as SS433 or M87.  Other radiatively inefficient models include Advection-Dominated Accretion Flows (ADAFs, Narayan \& Yi 1995) and  Magnetically-Dominated Accretion Flows (MDAFs, Fragile \& Meier 2009). These models, however, are more suitable for low accretion rates.

GX 339-4 was also detected during the low luminosity phase of the low-hard state in 1999 (Obs3 and Obs4). The typical X-ray luminosity is $L_X\approx10^{34}$ erg s$^{-1}$. Applying the same energetic considerations as above, the minimum jet power required is now $L_{\rm{jet}}\approx2\times10^{35}$ erg s$^{-1}$, a fraction $q_{\rm{jet}}\approx3\times10^{-4}$ of the Eddington luminosity of the black hole.

The observed spectrum in the X-ray band is quite hard, $L_X\propto E_\gamma^{-p}$ with $p\approx0.3$. If the X-rays originate in electron synchrotron radiation, from the observed slope it is possible to estimate the spectral index $\delta$ of the steady-state parent particle distribution, $N\propto E^{-\delta}$. They are related as

\begin{equation}
p=- \frac{\delta}{2}+\frac{3}{2}.
	\label{slopes1}
\end{equation}

\noindent This yields $\delta\approx 2.4$. Since particles cool, the index $\delta$ is not the same as that of the electron injection function, $Q\propto E^{-\alpha}$. In particular $\delta = \alpha+1$ in the case of dominant synchrotron losses. The particle injection spectrum must therefore be quite hard, with a power-law index smaller than the typically assumed $\alpha=2.0-2.2$ predicted by the theory of acceleration in strong, non-relativistic shocks. Here we decided to fix $\alpha=1.5$, consistent with relativistic shock acceleration (Stecker, Baring \& Summerling 2007). 

The values of  the relevant parameters of the model are summarized in Table \ref{model-parameters}. We allowed $q_{\rm{jet}}$, $a$, $\eta$, $E^{\rm{min}}$ and $\rho$ to vary subject to the constraints discussed above, whereas the rest of the parameters were kept fixed during the fitting process. We performed least-squares fits (Kay 1993) to the observational data. The quality of the fits was quantified calculating the correspondent value of $\chi^2$,

\begin{equation}
 \chi^2 = \sum \frac{(F_{\rm{obs}}- F_{\rm{m}})^2}{\Delta F_{\rm{obs}}^2}.
	\label{merit_factor}
\end{equation}

\noindent Here $F_{\rm{obs}}$ is the observed flux, $F_{\rm{m}}$ is the value predicted by the model, and $\Delta F_{\rm{obs}}$ is the uncertainty associated with every observational point. The best fit for a given set of parameters was found minimizing the value of $\chi^2$.

\begin{table}
\begin{center}
\caption{Values of the relevant parameters of the model.}
\label{model-parameters}
\begin{tabular}{@{}lcc}
\hline
Parameter & Symbol & Value\\
\hline 
Distance                                     & $d$                  & 6 kpc  \\ 
Black hole mass                              & $M_{\rm{BH}}$        & $6M_\odot$  \\
Viewing angle                                & $\phi$               & 30$^\circ$ \\
Jet bulk Lorentz factor                      & $\Gamma_{\rm{jet}}$  & 2  \\
Jet injection point                          & $z_0$          			& $4.5\times10^7$ cm  \\
Ratio $r_0/z_0$                              & $\chi$         			& 0.1  \\    
Ratio $2L_{\rm{jet}}/L_{\rm{Edd}}$ 					 & $q_{\rm{jet}}$ & $\ga10^{-4}$  \\           
Ratio $L_{\rm{rel}}/L_{\rm{jet}}$         	 & $q_{\rm{rel}}$ 			& 0.1  \\
Ratio $L_p/L_e$                           	 & $a$            		  & $\geq1$  \\
Magnetic field decay index $\quad\qquad$     & $m$                  & $1-2$  \\
Ratio $U_B/U_{\rm{(k,m)}}$ at $z_{\rm{acc}}$ & $\rho$               & $0.1-1$  \\
Particle injection index                 		 & $\alpha$             & $1.5$  \\
Minimum particle energy                   	 & $E^{\rm{min}}$       & $\geq2\,mc^2$ \\
Acceleration efficiency                      & $\eta$               & $10^{-4}-0.1$  \\
\hline
\end{tabular}
\end{center}
All parameters were kept fixed during fitting except $q_{\rm{jet}}$, $a$, $\eta$, $E^{\rm{min}}$ and $\rho$, that varied in the range indicated.\\
The value adopted for the jet Lorentz factor is typical for microquasar jets (e.g. Fender, Belloni \& Gallo 2004).
\end{table}

\subsection{Spectral energy distributions}

\begin{figure*}
\centering
\includegraphics[width=0.49\textwidth, keepaspectratio, trim=50 15 50 40, clip]{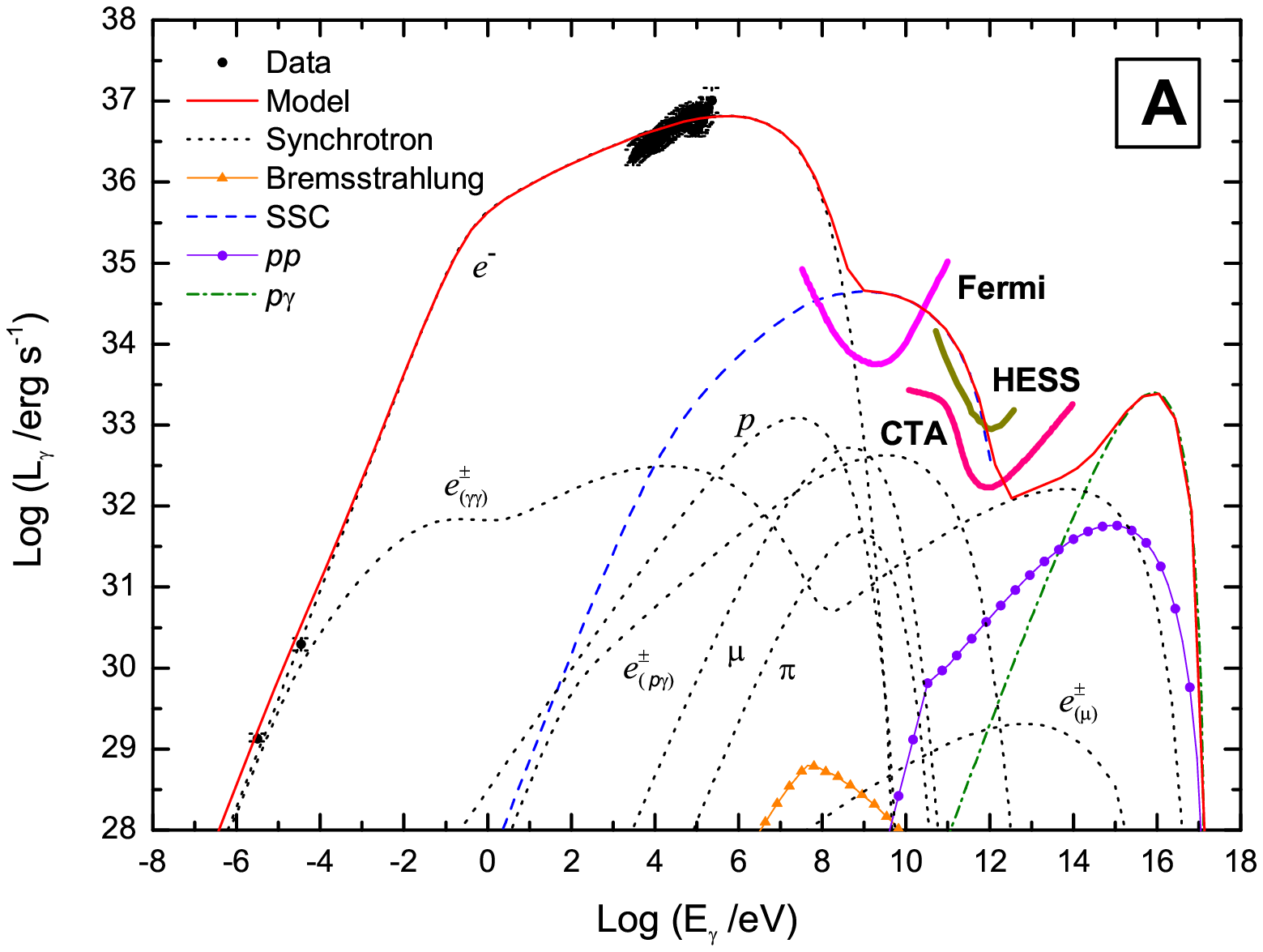}
\includegraphics[width=0.49\textwidth, keepaspectratio, trim=50 15 50 40, clip]{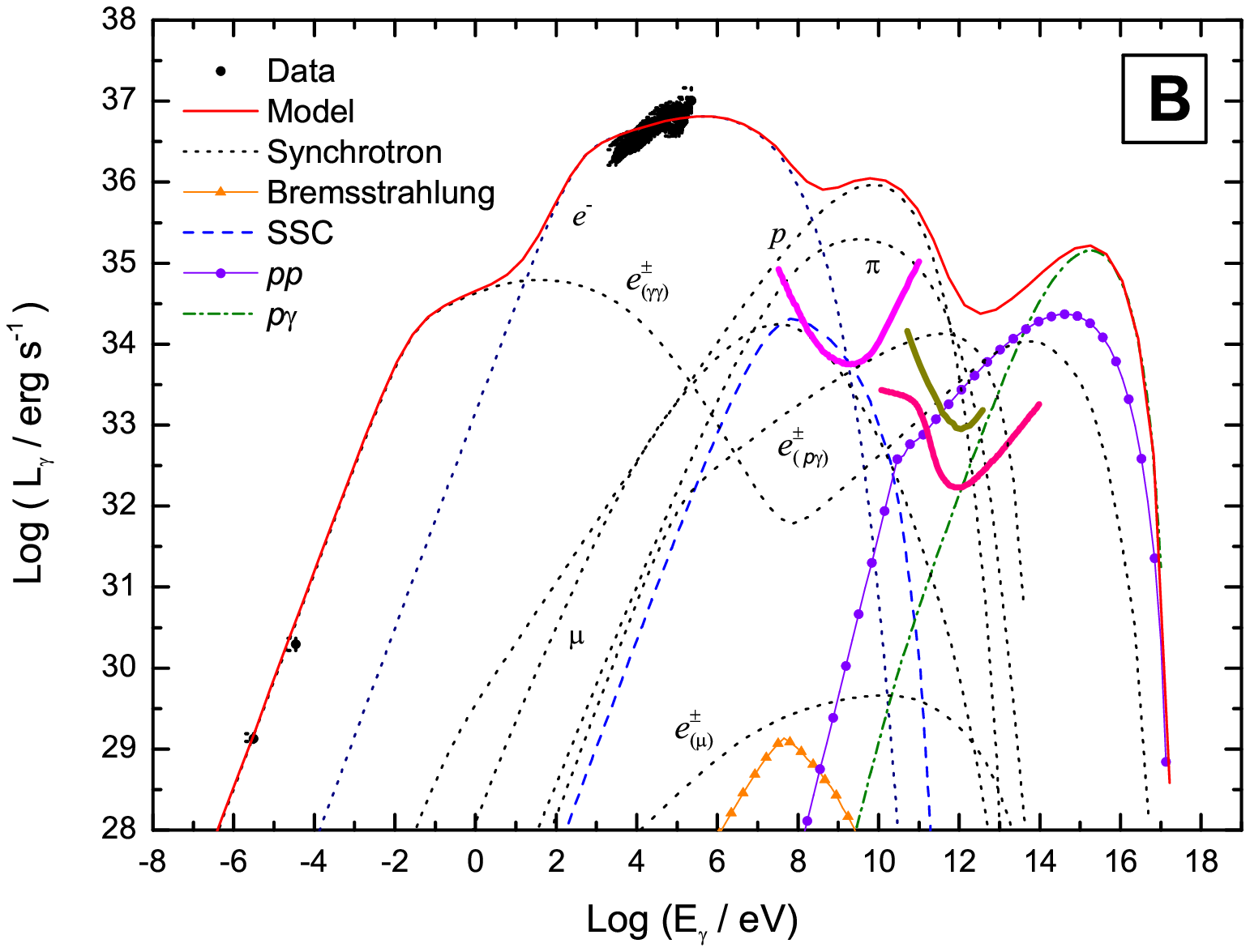}
\includegraphics[width=0.49\textwidth, keepaspectratio, trim=50 20 50 40, clip]{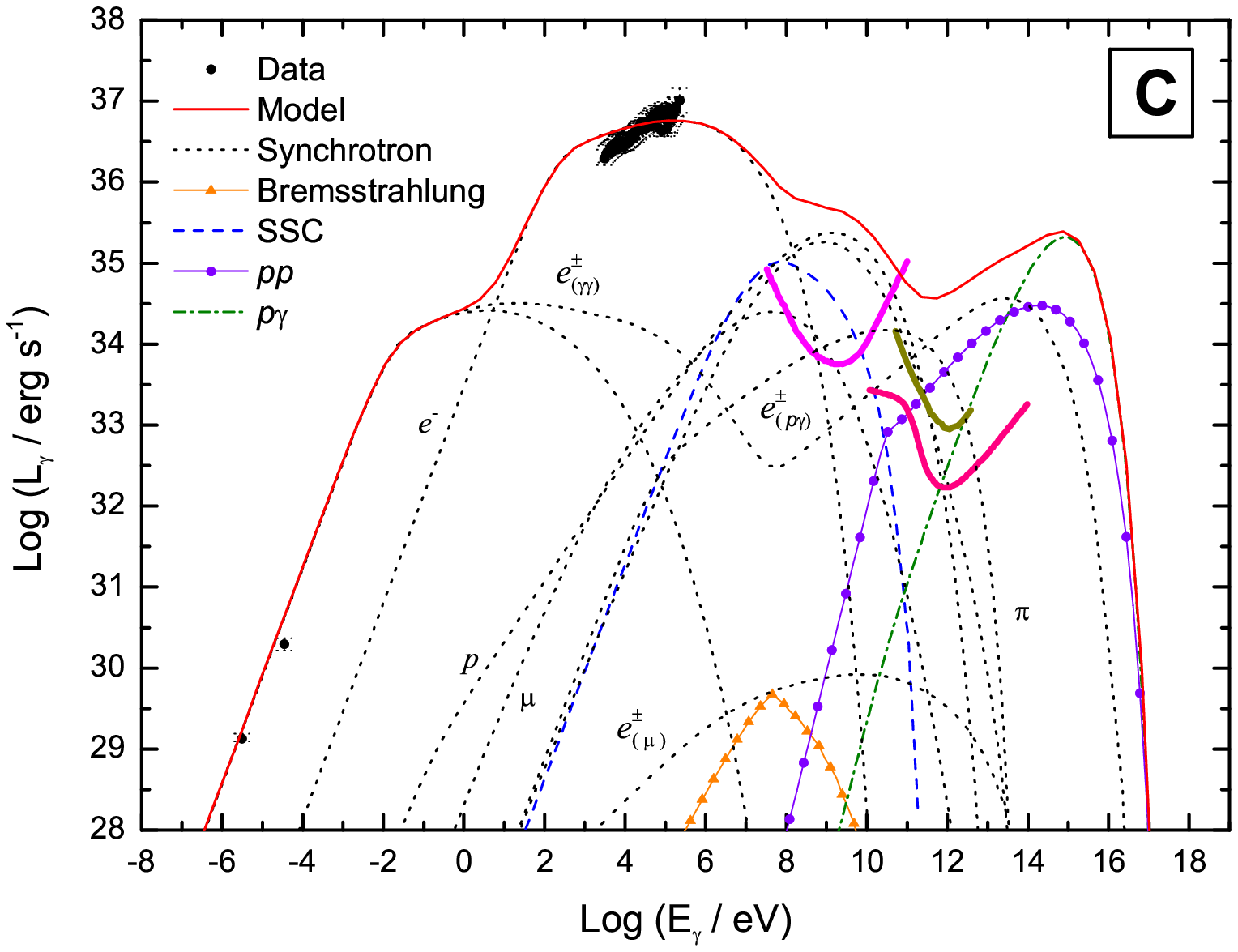}
\includegraphics[width=0.49\textwidth, keepaspectratio, trim=50 20 50 40, clip]{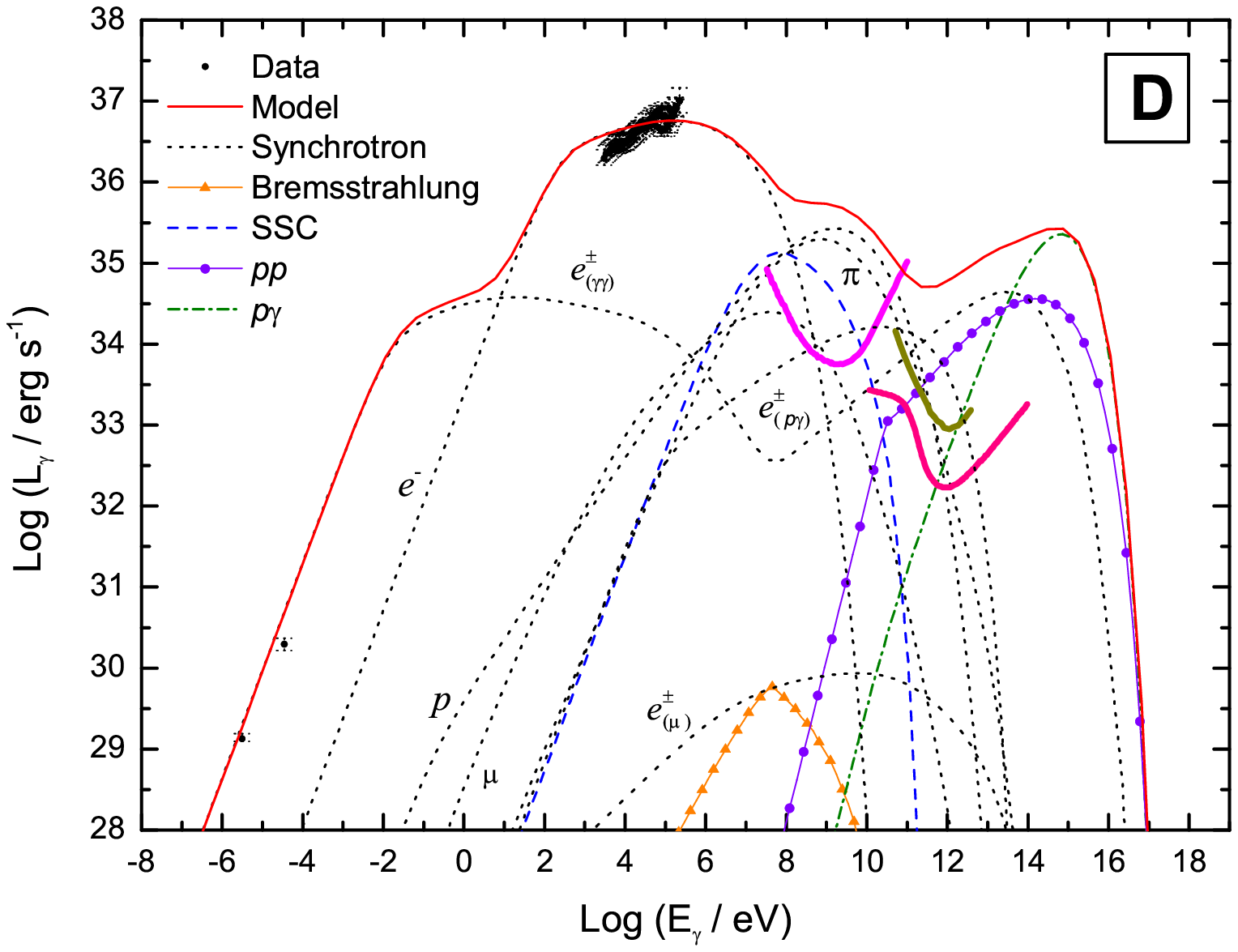}
\caption{Best fit spectral energy distributions (SEDs) of Obs1 for different values of the magnetic field decay index $m$. The graphics are labeled as in Table \ref{fit-parameters}. The position $z_{\rm{acc}}$ of the acceleration region was determined demanding that $U_{\rm{B}}<U_{\rm{k}}$. See Tables \ref{model-parameters} and \ref{fit-parameters} for the values of the rest of the parameters. The subindices $(\gamma\gamma)$, $(p\gamma)$ and $(\mu)$ indicate pairs created through photon-photon annihilation, photopair production and muon decay, respectively. The thick lines are the sensitivity limits of Fermi and HESS and CTA. }
\label{SEDs1}
\end{figure*}

\begin{figure*}
\centering
\includegraphics[width=0.49\textwidth, keepaspectratio, trim=50 15 50 40, clip]{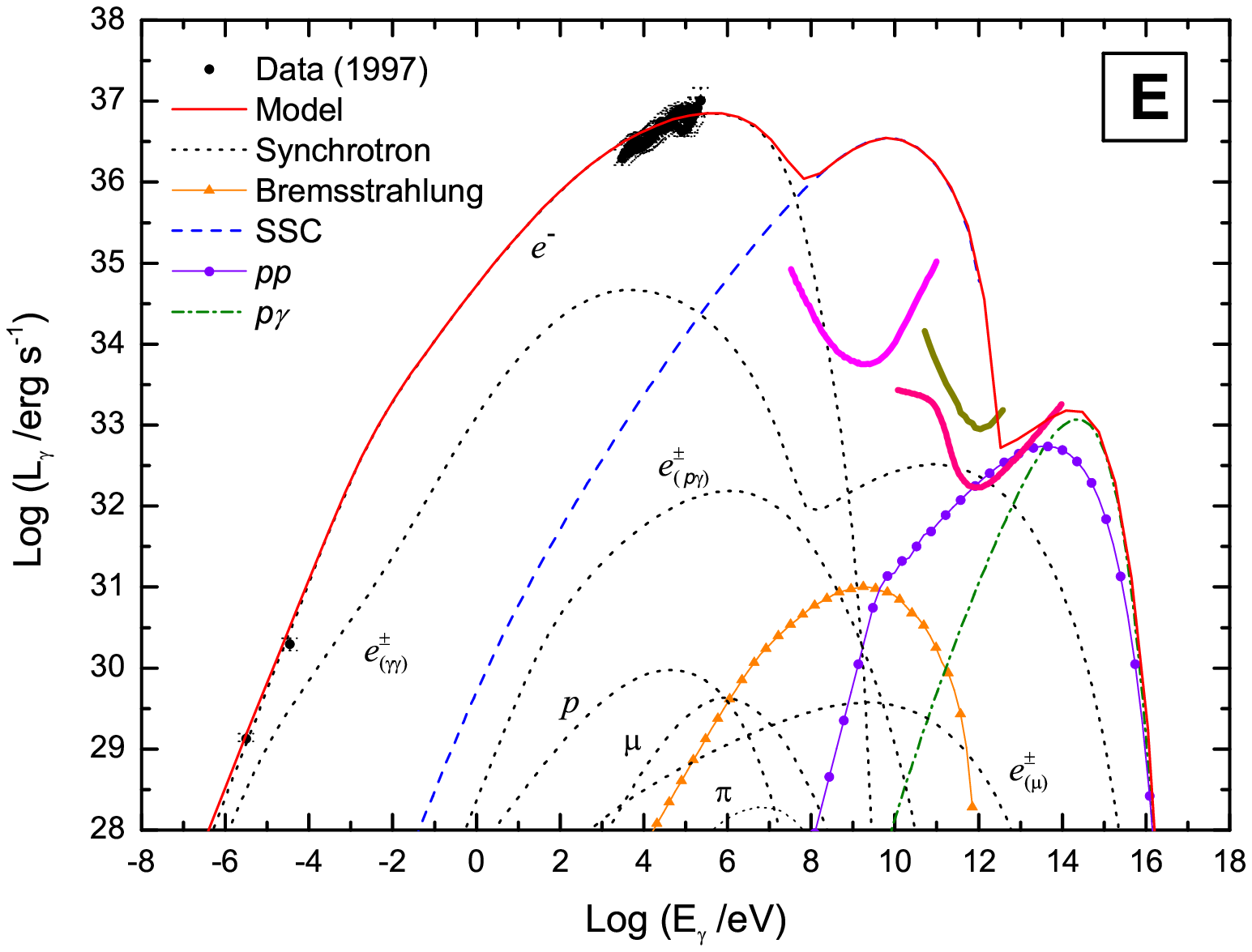}
\includegraphics[width=0.49\textwidth, keepaspectratio, trim=50 15 50 40, clip]{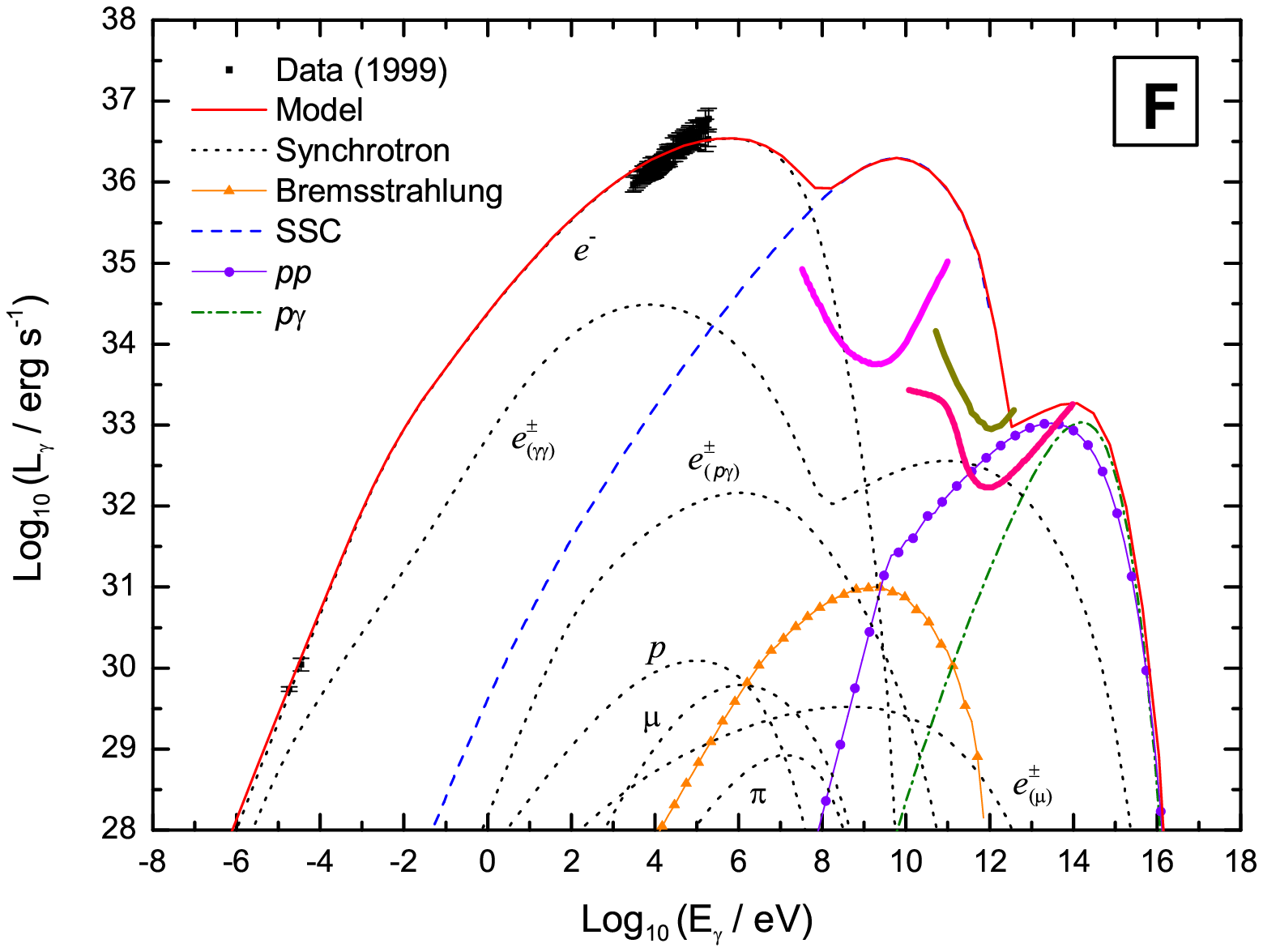}
\caption{The same as in Figure \ref{SEDs1}, but with $z_{\rm{acc}}$ calculated from the condition $U_{\rm{B}}<U_{\rm{m}}$. Model E is a fit to Obs1 and model F to Obs2.}
\label{SEDs2}
\end{figure*}

Figures \ref{SEDs1} and \ref{SEDs2} show the best fits obtained for a set of simultaneous radio and X-ray data taken on February 3rd 1997 and April 2nd 1999 (Obs1 and Obs2 in Table \ref{obs_data}, respectively), when the source was in a luminous low-hard state. In the case of Figure \ref{SEDs1}, the location of the acceleration region $z_{\rm{acc}}$ was determined through condition (\ref{sub-equipartition}), whereas in the case of Figure \ref{SEDs2} condition (\ref{sub-equipartition_2}) was applied. 

Each graphic corresponds to a different value of the magnetic field decay index $m$. This parameter strongly determines the shape of the spectrum, since it fixes the value of the field along the jet and consequently $z_{\rm{acc}}$. Larger values of $m$ yield $z_{\rm{acc}}$ closer to the jet base where the magnetic field is stronger. The values of the parameters result from the fitting are listed in Table \ref{fit-parameters}.

\begin{table*}
\begin{minipage}{150mm}
\caption{Best fit values of the relevant parameters.}
\label{fit-parameters}
\begin{tabular}{@{}ccccccccccccc}
\hline
Model     & Obs & $\alpha$ & $m$ &  $E_{\rm{min}}/mc^2$ & $q_{\rm{jet}}$ & $a$ & $\eta$ & $\rho$ & $z_{\rm{acc}}$ [$R_g$] & $B(z_{\rm{acc}})$ [G] & $\dot{N}_{e^+}$ [s$^{-1}$] & $\chi^2$/d.o.f.\\
\hline 

A & Obs1 & $1.5$ & $1.2$ & $97.4$ & $0.9$ & $1.5$ & $0.1$ & $0.1$ & $9.7\times10^4$ & $8.4\times10^{3}$  & $3.8\times10^{38}$ & $1.42$ \\ 

B & Obs1 & $1.5$ & $1.5$ & $99.2$ & $0.8$ & $1.4$ & $0.08$ & $0.75$ & $1.4\times10^2$ & $1.5\times10^{7}$  & $8.6\times10^{40}$ & $2.0$ \\

C & Obs1 & $1.5$ & $1.8$ & $96.3$ & $0.8$ & $1.6$ & $0.03$ & $0.5$ & $1.4\times10^2$ & $1.4\times10^{7}$  & $8.4\times10^{40}$ & $3.1$ \\

D & Obs1 & $1.5$ & $2.0$ & $92.4$ & $0.75$ & $1.4$ & $0.03$ & $0.75$ & $85.5$ & $2.3\times10^{7}$  & $1.3\times10^{41}$ & $3.3$ \\

E & Obs1 & $1.5$ & $1.8$ & $15.0$ & $1.00$ & $1.5$ & $0.1$ & $0.1$ & $1\times10^4$ & $5.0\times10^{3}$  & $4.5\times10^{40}$ & $0.8$ \\ 

F & Obs2 & $1.5$ & $2.0$ & $11.5$ & $0.73$ & $2.7$ & $0.1$ & $0.1$ & $3.5\times10^3$ & $1.2\times10^{4}$  & $3.0\times10^{40}$ & $0.98$ \\ 

G & Obs3 & $1.5$ & $2.0$ & $2.0$ & $6.4\times10^{-3}$ & $2.0$ & $3\times10^{-3}$ & $0.4$ & $3\times10^3$ & $3.6\times10^{3}$  & $1.4\times10^{35}$ & $0.15$ \\ 

H & Obs4 & $1.5$ & $2.0$ & $2.0$ & $6.3\times10^{-3}$ & $2.0$ & $1\times10^{-4}$ & $0.15$ & $4.9\times10^3$ & $1.4\times10^{3}$  & $7.5\times10^{34}$ & $0.15$ \\ 

I & Obs3 & $2.2$ & $2.0$ & $25.2$ & $6.6\times10^{-3}$ & $2.0$ & $0.1$ & $0.4$ & $2\times10^2$ & $3.1\times10^{3}$  & $2.4\times10^{37}$ & $0.8$ \\ 

J & Obs5 & $1.5$ & $2.0$ & $92.7$ & $1.0$ & $1.0$ & $0.1$ & $0.1$ & $6\times10^3$ & $4\times10^{3}$  & $1\times10^{42}$ & $6.2$ \\ 

\hline
\end{tabular}
\\The values of $q_{\rm{jet}}$, $a$, $\eta$, $E^{\rm{min}}$ and $\rho$ were left free during the fitting procedure; $\chi^2$/d.o.f. is the value of $\chi^2$ per degree of freedom. 
\end{minipage}
\end{table*}

The spectral energy distributions (SEDs) presented in Figure \ref{SEDs1} all correspond to Obs1. The best fit is obtained for  $m=1.2$. The X-ray data range is always covered by the synchrotron emission of primary electrons, but as $m$ grows, synchrotron radiation of secondary pairs begins to dominate at radio wavelengths. This diminishes the quality of the fit. Also as $m$ increases the slope of the X-ray spectrum gets worse modeled, indicating that the injection index should be harder than the assumed $\alpha=1.5$. 

Between $\sim1$ GeV and $\sim1$ TeV, the emission is dominated by synchrotron self-Compton (SSC) radiation and synchrotron emission of protons and secondary particles; at higher energies the contributions of proton-proton and proton-photon interactions are the dominant ones. All these processes become more relevant when $z_{\rm{acc}}$ is nearer the jet base, since the magnetic field is stronger and enhances the synchrotron radiation of $p$, $\pi^\pm$ and $\mu^\pm$. Also the matter and photon densities are larger, providing denser targets for $pp$ and $p\gamma$ collisions, and SSC scattering. The contribution of secondary pairs from $\gamma\gamma$ annihilation is significantly increased for large values of $m$ due to this effect as well. In all cases the best fits favour large minimum particle energies, $E_{\rm{min}}\approx100\mathrm{mc}^2$. A powerful jet ($q_{\rm{jet}}\approx 0.8-0.9$) and equipartition of energy between primary protons and leptons ($a\approx1$) is also required, since the power injected in electrons needs to be as large as possible to account for the X-ray observations. 

Models E and F in Figure \ref{SEDs2} correspond to fits of Obs1 and Obs2, respectively. In both cases, $z_{\rm{acc}}$ was calculated demanding that $U_{\rm{B}}<U_{\rm{m}}$. For the same $m$ and $\rho$, this condition gives larger values of $z_{\rm{acc}}$ and weaker magnetic fields. Now the best fits are obtained for large values of $m$. These sets of parameters reproduce the slope of the X-ray spectrum for the same value of the injection index better than the models of Figure \ref{SEDs1}.

\begin{figure*}
\centering
\includegraphics[width=0.48\textwidth, keepaspectratio, trim=50 15 60 40, clip]{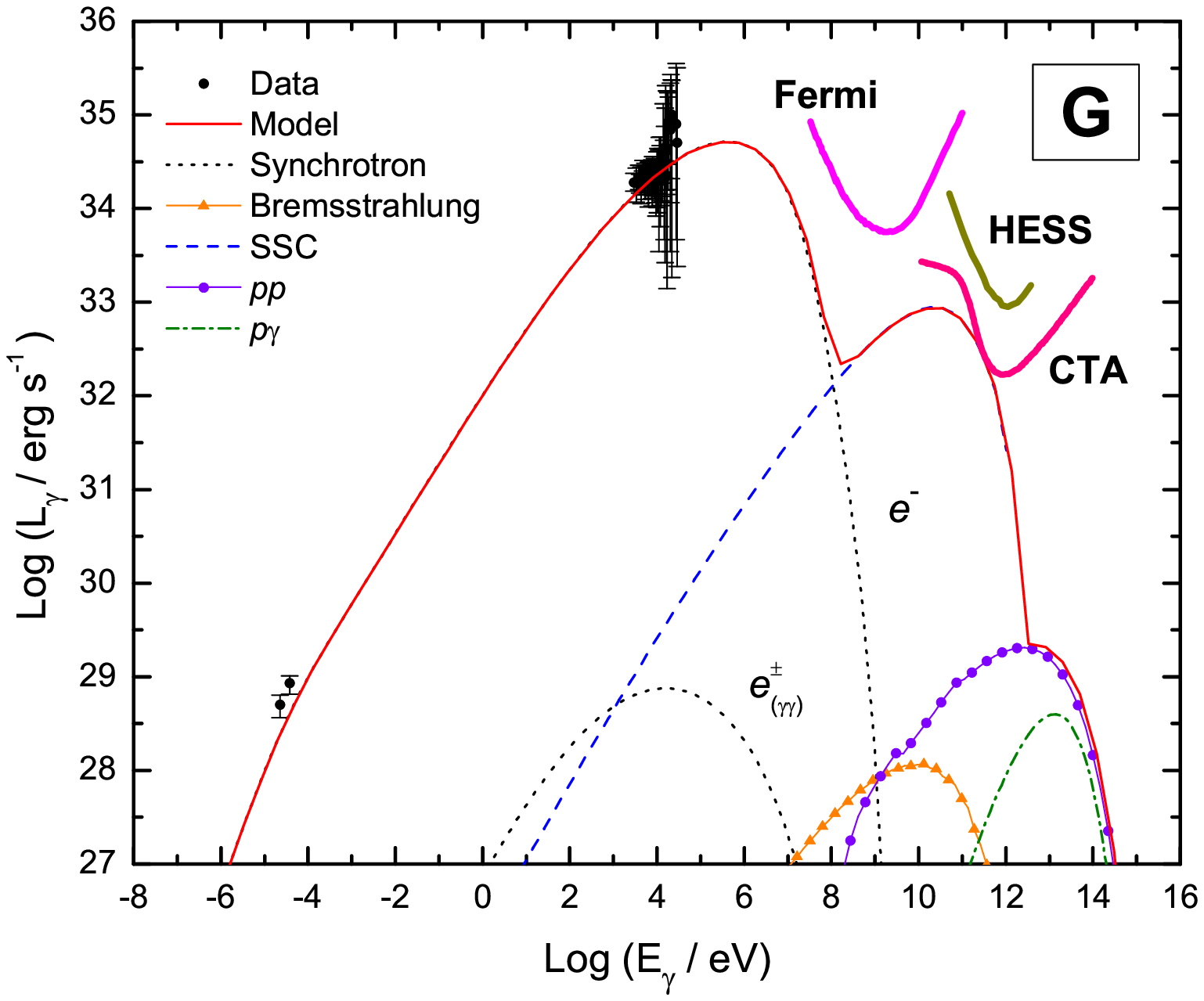}
\includegraphics[width=0.48\textwidth, keepaspectratio, trim=50 15 60 40, clip]{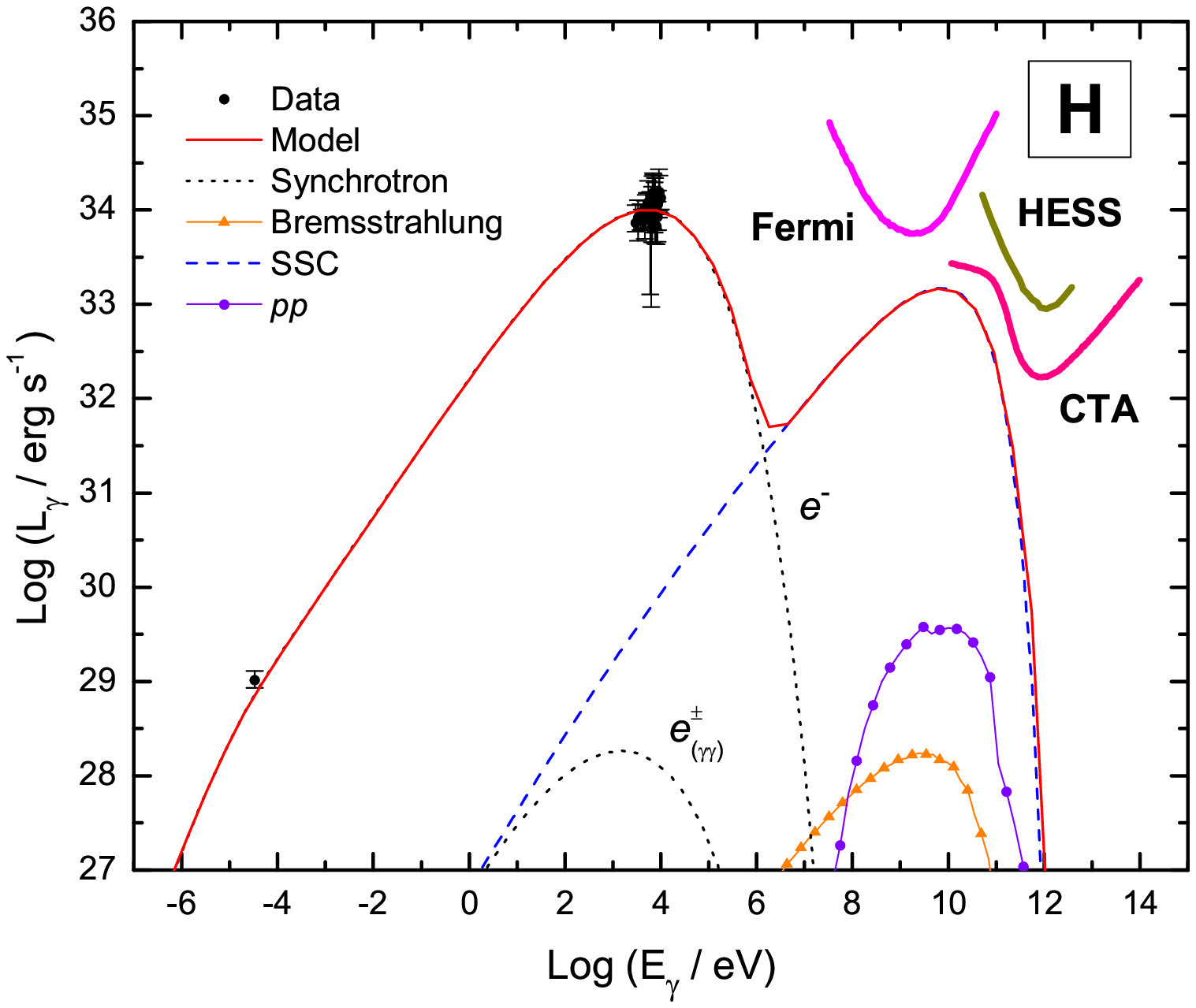}
\caption{Best-fitting SEDs of Obs3 (model G) and Obs4 (model H). The decay index of the magnetic field is $m=2$, and the position $z_{\rm{acc}}$ of the acceleration region was determined demanding that $U_{\rm{B}}<U_{\rm{m}}$. See Tables \ref{model-parameters} and \ref{fit-parameters} for the values of the rest of the parameters. The sensitivity limits of Fermi, HESS and CTA are indicated. Optical data in model G (not shown, see Figure \ref{SEDs4}) were not included in the fit.}
\label{SEDs3}
\end{figure*} 

Figure \ref{SEDs3} shows two model fits to low-luminosity low-hard state observations of GX 339-4, carried out in 1999 (Obs3 and Obs4).  The radio and X-ray emission is due to primary electrons; all radiative contributions of protons and secondary particles are negligible. The jet power required to account for the data is now only a fraction $q_{\rm{jet}}\approx6\times10^{-3}$ of the Eddington luminosity. The best fit models are obtained for low values of the acceleration efficiency and minimum particle energy.

For each model we calculated the synchrotron emission of thermal electrons at the base of the jet. For an electron energy  $E_e\approx2m_ec^2$ and a magnetic field $B_0\approx10^{6-7}$ G, the peak of the spectrum is at $E_\gamma\approx10$ eV. However, the luminosity of this component is below or just above the jet emission. This contribution is not significant in the relevant energy bands. Therefore, the results of the fits are not affected.
 
\subsection{Radio/X-rays and NIR/X-rays correlations}

The analysis of simultaneous radio and X-rays observations from 1997-1999, led Corbel et al. (2003) to find out that the fluxes in both energy bands are tightly correlated. They showed that the radio flux at 8.6 MHz is related the 3-9 keV integrated X-ray flux as \mbox{$F_{\rm{R}}\propto \Delta F_{\rm{X}}^{\delta}$}. This correlation suggests a common origin in the jet\footnote{Alternative models to explain the radio/X-rays correlation have been suggested. Markoff et al. (2005)  presented fits to simultaneous radio and X-ray data of GX 339-4 obtained applying a corona model. Furthermore, in Heinz \& Sunyaev (2003) it is shown that for an ADAF-like boundary condition, the radio flux from the base of the jet scales with the black hole mass and the accretion rate, independently of the assumed jet model.} (synchrotron radiation). According to Markoff et al. (2003), if all model parameters except the jet power are kept frozen, the correlation index $\delta$ is given by

\begin{equation}
 \delta = \frac{17/12-2/3\delta_{\rm{R}}}{17/12-2/3\delta_{\rm{X}}}.
	\label{corr-index}
\end{equation}

\noindent Here $\delta_{\rm{X}}$ is the spectral index of the X-ray region of the synchrotron spectrum $(F_{\rm{X}}\propto \nu^{\delta_{\rm{X}}})$, and $\delta_{\rm{R}}$ that of the synchrotron radio  flux $(F_{\rm{R}}\propto \nu^{\delta_{\rm{R}}})$.

\begin{figure}
\centering
\includegraphics[width=\columnwidth, keepaspectratio, trim=40 15 60 40, clip]{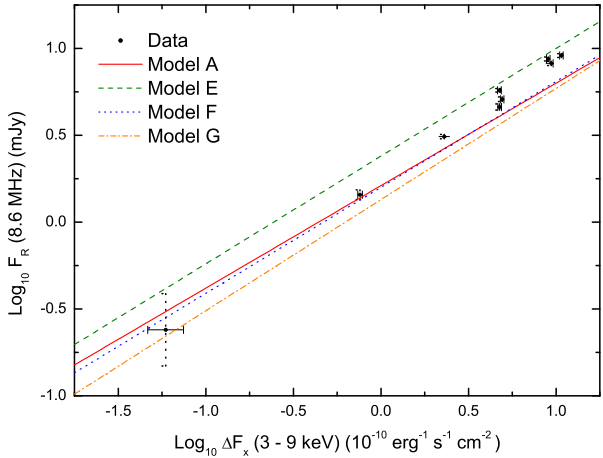}
\caption{Radio/X-ray flux correlations in GX 339-4. The different curves correspond to models A, E, F and G. In each case, the slope was calculated as in Markoff et al. (2003), and to determine the intercept we used corresponding SEDs in Figures \ref{SEDs1}, \ref{SEDs2} and \ref{SEDs3}. The model correlation index is $\delta\sim 0.6$, whereas that of the 1997-1999 observational data is $\delta\sim0.7$ (Corbel et al. 2003).}
\label{correlations}
\end{figure}

In our model radio and X-ray emission is due to synchrotron radiation of electrons. We find a value of the radio spectral index $\delta_{\rm{R}}\sim 0.33$, which corresponds to the optically thick part of the spectrum from a particle distribution with a low-energy cutoff. The value of the X-ray spectral index is $\delta_{\rm{X}}\sim -0.8$, as expected for an injection function $Q_e\propto E^{-1.5}$ (notice that electrons are then strongly cooled due to synchrotron losses). These values yield $\delta\sim 0.6$. Figure \ref{correlations} shows the correlation curves predicted by our model for cases A, E, F and G, together with the data from Corbel et al. (2003). The model results are in reasonable agreement with the observations.

Simultaneously with the radio and X-ray observations of Obs3 and Obs5, GX 339-4 was also detected at NIR and optical wavelengths (Markoff et al. 2003; Homan et al. 2005). As in the case of the radio emission, the NIR/optical flux also displays a strong correlation with the X-ray flux. 

From an analysis of low-hard state data from the 2002 outburst of GX 339-4 (same epoch as Obs5), Homan et al. (2005) shown that the flux density in the NIR \emph{H}-band and the $3-100$ keV bolometric X-ray flux correlate as $F_H\propto \Delta F_{\rm{X}}^\delta$, with $\delta=0.53$. A similar correlation was found between the optical \emph{V}-band and \emph{I}-band flux densities and the integrated X-ray flux, with correlation indices $\delta=0.44$ and $\delta=0.48$, respectively. These correlations disappear when the source leaves the low-hard state. The \emph{H}-band emission, however, rises and decays faster than the optical during the state transition, while the slope between the \emph{I} and \emph{V} bands remains constant. As suggested by Homan et al. (2005), this may indicate a different origin for the NIR and optical emission during the low-hard state.

The correlations between the radio/X-ray and NIR/X-ray fluxes suggest that the emission in the three ranges must originate in the jet. This is further supported by the fact that the NIR flux extrapolates back to the radio data (see also Corbel et al. 2003). Direct or reprocessed emission from an accretion disc can be ruled out due to the shape of the NIR/optical spectrum and the short decay time scales. Furthermore, the NIR and radio fluxes are quenched when the disc begins to contribute significantly to the X-ray emission. Homan et al. (2005) conclude that the NIR emission probably originates in the jet, and approximately coincides with the position of the break of the synchrotron spectrum. The optical flux may be due to thermal and/or non-thermal reprocessed radiation from the accretion disc or star, or from a region of the jet different from where the NIR emission is produced.

These ideas are supported by the recent results of Coriat et al. (2009), who presented an analysis of five years of observations of GX 339-4 (from 2002 to 2007, a period that comprises five outbursts). They found a strong IR/X-ray correlation over four decades in flux during the low-hard state. The correlation index, however, is not unique: a break appears at bolometric (3-100 keV) X-ray fluxes $\sim1.1\times10^{-10}$ erg s$^{-1}$ cm$^{-2}$  ($L_X\sim6\times 10^{-4}L_{\rm{Edd}}$ for $M=6M_\odot$ and $d=6$ kpc). Coriat et al. (2009) argue that this break can be explained attributing the X-ray emission to SSC radiation from the jet (see also Nowak et al. 2005, where it is suggested that models more complex than a single jet synchrotron component maybe needed to explain the correlations). They find no clear evidence of a similar break in the \emph{V}-band/X-ray correlation, and suggest that the optical emission in the low-hard state is dominated by the outer part of the accretion disc, and not by the jet. 

This correlation is not peculiar of GX 339-4, but it seems to be a signature of low-mass black hole X-ray binaries. Russell et al. (2006) analysed radio, NIR, optical and X-ray data from 16 sources (including extragalactic systems in the LMC). Their results agree with those of Homan et al. (2005) for GX 339-4. They estimate that the jet contribution to the NIR emission during the low-hard state is $\sim90\%$, but only $\sim50\%$ to the \emph{I} and \emph{V} bands. 

\begin{figure*}
\centering
\includegraphics[width=0.45\textwidth, keepaspectratio, trim=50 15 60 40, clip]{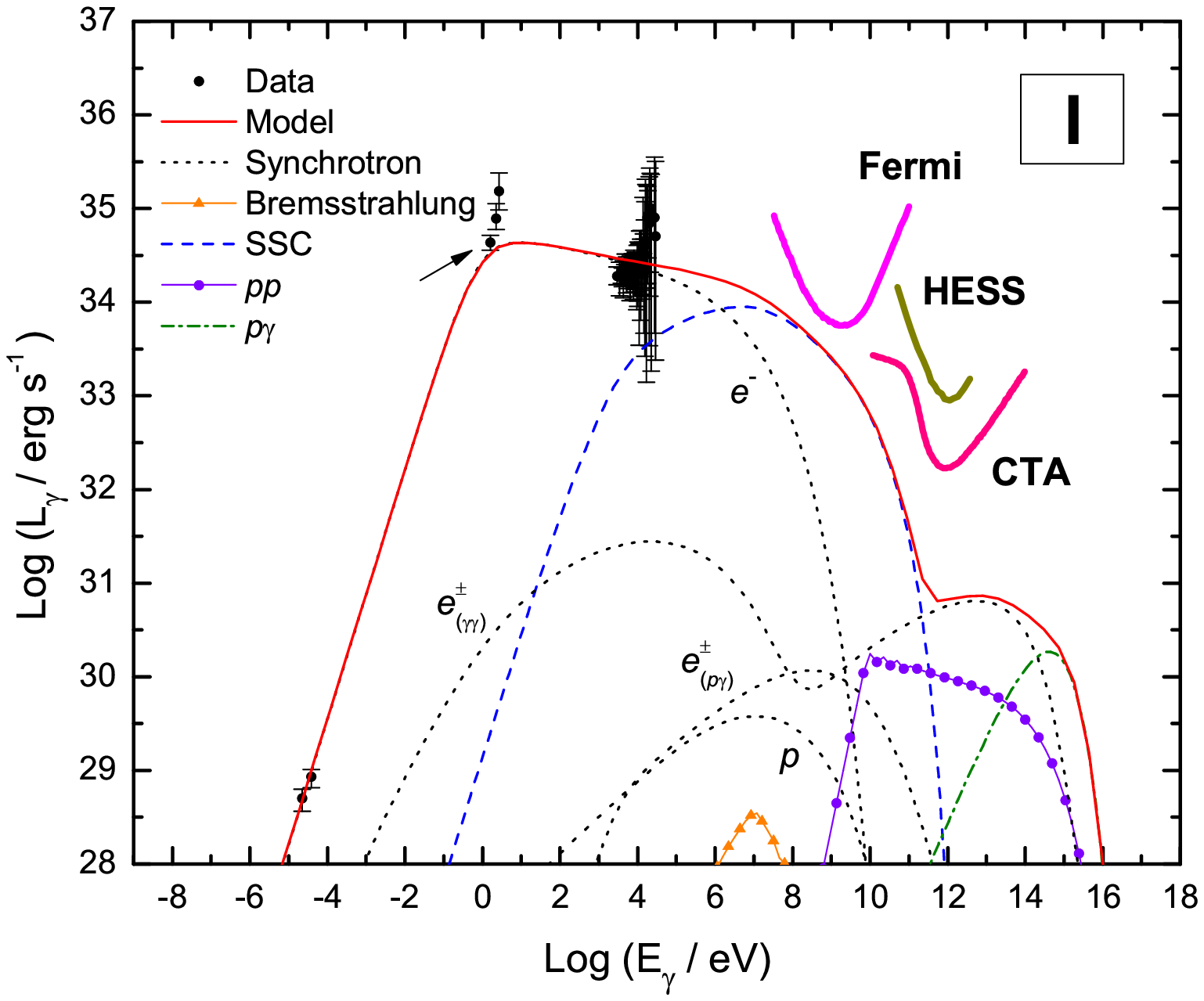}
\includegraphics[width=0.48\textwidth, keepaspectratio, trim=50 15 73 40, clip]{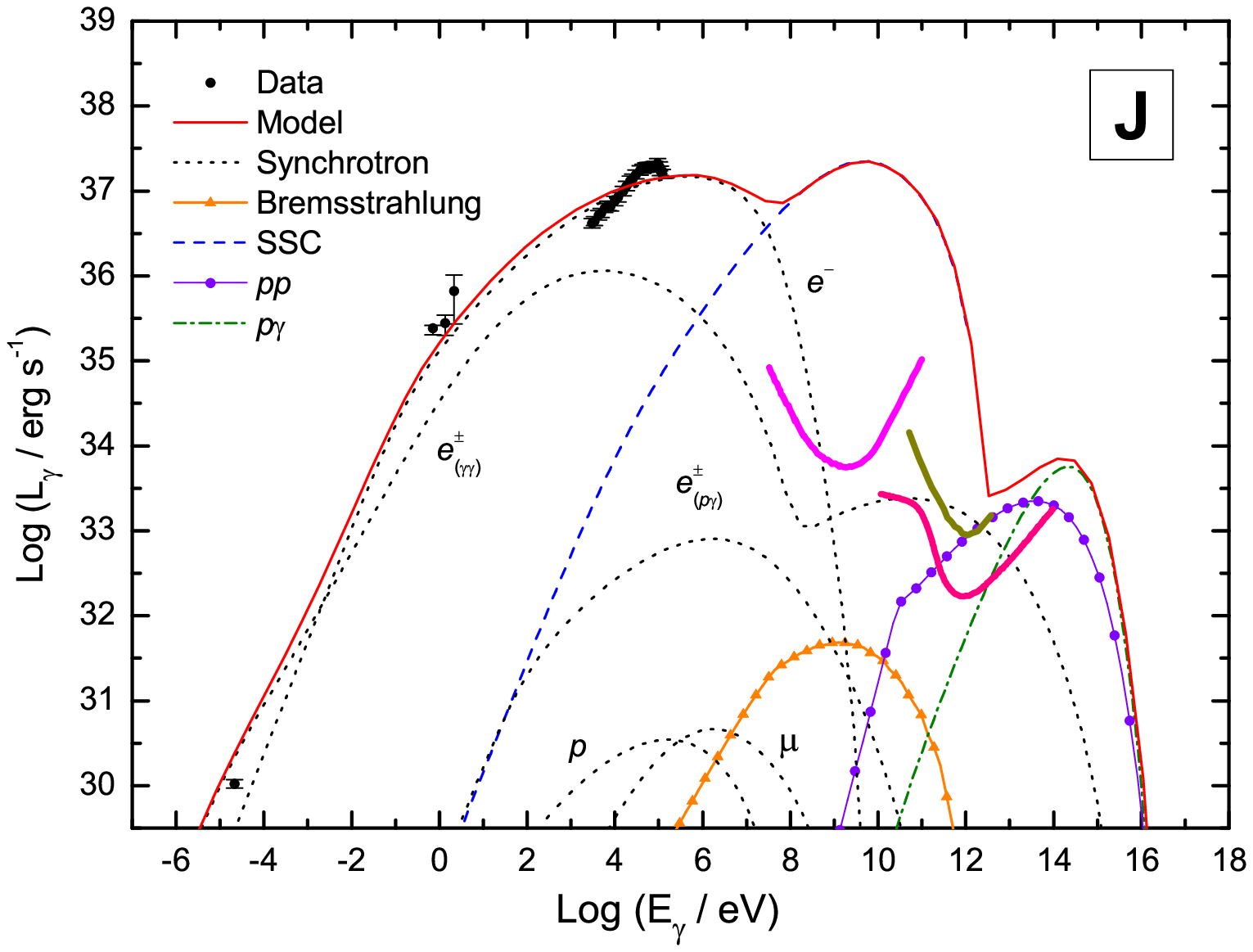}
\caption{Best fit SEDs of Obs3 (model I) and Obs5 (model J). In model I,  the arrow indicates the only point in the optical band that was included in the fit. The decay index of the magnetic field is $m=2$. The position $z_{\rm{acc}}$ of the acceleration region was determined demanding that $U_{\rm{B}}<U_{\rm{k}}$ in Model I, and  $U_{\rm{B}}<U_{\rm{m}}$ in Model J. See Tables \ref{model-parameters} and \ref{fit-parameters} for the values of the rest of the parameters. The sensitivity limits of Fermi, HESS and CTA are indicated. }
\label{SEDs4}
\end{figure*} 

Figure \ref{SEDs4} shows the best fit models obtained for Obs3 (including the lowest energy point in the optical) and Obs5. In the case of Obs5, the radio, NIR, optical and X-ray data are reasonably well reproduced using a hard particle injection spectral index $\alpha=1.5$. In the case of Obs3, the data at optical frequencies cannot be accounted for with a single synchrotron component. However, using a softer particle injection spectral index $\alpha=2.2$, it is possible to obtain models models where the synchrotron turnover occurs in the optical\footnote{Synchrotron radiation of thermal electrons from the base of the jet is not relevant in these models either.}. The rise in the spectrum at higher energies, however, cannot be fitted. This emission must have a different origin, for example in an accretion disc (Markoff et al. 2003).

\subsection{Photon absorption}

\begin{figure}
\centering
\includegraphics[width=\columnwidth, keepaspectratio, trim=40 15 60 40, clip]{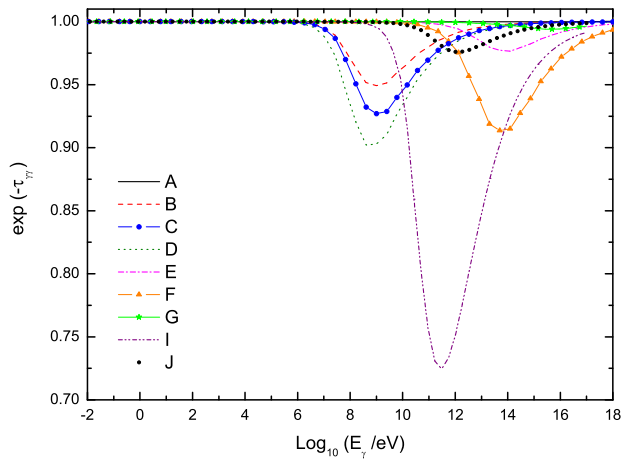}
\caption{Attenuation factor for a photon of energy $E_\gamma$ due to $\gamma\gamma$ annihilation in the total photon field of the emission region. Absorption is negligible in all cases and does not modify the production spectrum.}
\label{attenuation}
\end{figure}

In order to asses the effect of photon self-absorption by $\gamma\gamma$ annihilation, we calculated the attenuation parameter $\exp(-\tau_{\gamma\gamma})$ in equation (\ref{attenuated-luminosity}). As it can be seen from Figure \ref{attenuation}, internal attenuation is almost negligible due to the low photon density in the emission region. The production spectrum is not appreciably modified in any case, contrary to some models calculated by Romero \& Vila (2008).

\subsection{Discussion}

According to the results of Figures \ref{SEDs1}, \ref{SEDs2} and \ref{SEDs4}, during the high luminosity low-hard state, GX 339-4 would be a source detectable by Fermi in the energy range 100 MeV - 10 GeV. In this range the emission is due to electron synchrotron radiation and SSC. However, it will be far from being a bright source, possibly appearing with a significance of 5-6 $\sigma$ after one year of exposure. At higher energies, the emission of hadronic origin will be harder to detect with the present instruments. For some models, the predicted luminosities at $\sim1$ TeV are above the sensitivity of HESS for a point source at $6$ kpc through exposures of more than 50 hours. Future instruments like the Cherenkov Telescope Array (CTA) will make it easier to obtain a firm detection. Such a detection at high energies is crucial to evaluate the hadronic content of the jet.  
On the other hand, when the source is in a low luminosity phase, according to the results in Figure \ref{SEDs3} it would be undetectable at high energies with the present instruments.

\section{Positron production rate}
\label{positrons}

Recently, an analysis of several years of data collected by the INTEGRAL satellite led Weidenspointner et al. (2008) to suggest that the observed asymmetry in the distribution of the electron-positron annihilation line traces the distribution of hard low-mass X-ray binaries (LMXBs) in the Galaxy. These authors discovered that the 511 keV line emission from the inner galactic disc is clearly asymmetric: the flux from the region of negative galactic longitudes is 1.8 times larger than that from the corresponding region of positive longitudes. The same type of asymmetry is observed in the spatial distribution of hard LMXBs, those that show appreciable emission above 20 keV. The authors argue that positrons escaping the injection region with an energy of $\sim1$ MeV will not diffuse more than $\sim100$ pc before annihilating. This distance is short enough for the line emission to be still correlated with the large-scale distribution of the sources. Weidenspointner et al. (2008) estimate that the positron production rate required to account for the observed flux is $\sim10^{41}$ erg s$^{-1}$. 

To investigate this possibility we calculated the positron injection rate in our model. According to Heinz (2008), the number of injected positrons per unit time can be roughly estimated as

\begin{equation}
\dot{N}_{e^+}\approx \frac{L_{e^\pm}}{2\Gamma_{\rm{jet}}\bar{\gamma}m_ec^2}.
	\label{positron_injection_rate}
\end{equation} 

\noindent In this expression $L_{e^\pm}$ is the total luminosity injected in pairs and $\bar{\gamma}$ is the mean Lorentz factor of the positrons when they leave the source. 

It is reasonable to expect that positrons have almost completely cooled when they reach the end of the jet, and thus $\bar{\gamma}$ is of the order of the jet bulk Lorentz factor, $\bar{\gamma}\sim\Gamma_{\rm{jet}}=2$. In our case, the most relevant process of pair production is $\gamma\gamma$ annihilation. The calculated positron injection rates are shown in Table \ref{fit-parameters}. For those models that correspond to the high X-ray luminosity states, $\dot{N}_{e^+}$ is very near or even exceeds the necessary lower limit given by Weidenspointner et al. (2008). There are about $\ga100$ LMXB in the Galaxy and, although not detected yet, possibly most of them produce jets. Even if many of them are considerably less powerful than the jet in GX 339-4, the added contributions might well account for the observed flux at 511 keV. Our estimations show that the proposed association between LMXB and the  annihilation line emission is indeed feasible at least in energetic terms. In this way, there may be no need to resort to other more exotic explanations, such as annihilation of dark matter. A more detailed discussion on this topic will be presented elsewhere.

\section{Summary and conclusions}
\label{Conclusions}

We have shown that, under certain general conditions, the model developed here is capable of explaining the observed radio and X-ray spectrum of the low-mass microquasar GX 339-4. The parameter that mainly determines the characteristics of the obtained spectra is the magnetic field decay index $m$. This parameter, in turn, depends on unknown factors as the geometry and turbulence level of the magnetic field in the inner jet. The best fits to the high luminosity 1997 and 1999 data are obtained in those cases where the acceleration/emission region is placed relatively far from the compact object and the jet apex, $z_{\rm{acc}}\sim10^{9}$ and $z_{\rm{acc}}\sim10^{11}$ cm (models A and E, respectively). In both cases the value of the magnetic field is similar, $B(z_{\rm{acc}})\sim10^4$ G, although $m=1.2$ in model A and $m=2$ in model E. The slope of the X-ray spectrum was difficult to reproduce, even assuming a hard particle injection with spectral index $\alpha=1.5$. Nevertheless, a harder particle injection could be achieved in principle through diffusive acceleration at relativistic shocks, see for example the works of Stecker, Baring \& Summerling (2007) and Keshet \& Waxman (2005). In any case, the high X-ray luminosities require a powerful jet with a large leptonic content. In fact, all the fits yield $a\sim1$, what means that as much energy is given to the primary relativistic electrons as it is allowed by the constraints imposed. The hadronic contribution to the spectrum in cases A and E is undetectable with the present gamma-ray instruments. In the other models, synchrotron radiation of protons and secondary muons and pions, and at higher energies the contribution of $pp$ interactions, could be detectable by Fermi and HESS (and in the future by CTA), respectively. 

For the low-luminosity observations (models G and H), the best fits were obtained for low values of the acceleration efficiency and minimum particle energy. The required jet luminosity is now only $L_{\rm{jet}}\sim6\times10^{-3}L_{\rm{Edd}}$. The predicted emission above $\sim100$ MeV is too faint to be detected with the present gamma-ray telescopes.

We have also calculated fits to simultaneous radio, NIR/optical and X-ray observations from 1999 and 2002 (models I and J). For these sets of parameters, the break in the synchrotron spectrum occurs approximately in the NIR, and the lowest-energy data was reasonably fit. The rising shape of the spectrum at optical wavelengths, however, could not be reproduced. This component is likely to originate mostly outside the jet, probably in the accretion disc. 

In all models, the spectrum is essentially of leptonic origin. In this sense, the results do not differ from those of previous works like those of Markoff et al. (2003, 2005).  Our model, however, besides making predictions for the emission in the very high-energy regime, introduces some refinements over the previous scenarios adopted for this source. The particle distributions are calculated self-consistently taken into account the effect of energy losses on the injection spectrum. We also calculate the radiation emitted by secondary particles produced in hadronic interactions, and that of the electron-positron pairs product of photon-photon annihilation. The importance of photon self-absorption is assessed as well, although it turns out not to be relevant since the emission region is in a zone of low radiation density.

We have also shown that the pair injection rate is significant enough, if this kind of model is solid in general for low-mass microquasars, to account for the observed line emission at 511 keV, according to the lower limit given by Weidenspointner et al. (2008). If the proposed association between hard low-mass X-ray binaries and the electron-positron annihilation line flux can be proved, other explanations such as annihilation of dark matter could result unnecessary. Detailed models for dealing with this topic are in preparation.      

\section*{Acknowledgments}

G.S.V. thanks Nicol\'as Casco for his help with fitting algorithms and Mat\'ias Reynoso for useful discussion. We thank an anonymous referee for very useful comments. G.E.R. thanks Sera Markoff for insightful remarks and suggestions. This research was supported by ANPCyT through grant PICT-2007-00848 BID 1728/OC-AR and by the Ministerio de Educaci\'on y Ciencia (Spain) under grant AYA 2007-68034-C03-01, FEDER funds.

\label{lastpage}


\begin{thebibliography}{99}

\bibitem[\protect\citeauthoryear{AharonianBook}{2004}]{AharonianBook} Aharonian F.A., 2004, Very High Energy Cosmic Gamma Radiation. World Scientific Publishing, Singapore 
\bibitem[\protect\citeauthoryear{Atoyan}{2003}]{Atoyan03} Atoyan A.M., Dermer, C.D., 2003, ApJ, 586, 79
\bibitem[\protect\citeauthoryear{Blandford}{1999}]{Blandford99} Blandford R.D., Begelman M.C., 1999, MNRAS, 303, L1-L5
\bibitem[\protect\citeauthoryear{BG}{1970}]{BG70}Blumenthal G.R., Gould R.J., 1970, Rev. Mod. Phys., 42, 237
\bibitem[\protect\citeauthoryear{Bogovalov}{2005}]{Bogobalov05} Bogovalov S.V., Kelner S.R., 2005, Astronomy Reports, 49, 57
\bibitem[\protect\citeauthoryear{BR}{2006}]{BoschRamon06}Bosch-Ramon V., Romero G.E., Paredes J.M., 2006, A\&A, 447, 263
\bibitem[\protect\citeauthoryear{BS}{1997}]{BS97}B\"ottcher M., Schlickeiser R., 1997, A\&A, 325, 866
\bibitem[\protect\citeauthoryear{Buxton}{2003}]{Bux03}Buxton M., Vennes S., 2003, MNRAS, 342, 105
\bibitem[\protect\citeauthoryear{Callanan}{1992}]{Callanan92}Callanan P.J., Charles P.A., Honey W.B., Thorstensen J.R., 1992, MNRAS, 259, 395
\bibitem[\protect\citeauthoryear{Chodorowski}{1992}]{Ch92} Chodorowski M.J., Zdziarski A.A., Sikora M., 1992, ApJ. 400, 181
\bibitem[\protect\citeauthoryear{Corbel}{2000}]{Corbel00}Corbel S., Fender R.P., Tzioumis A.K., Nowak M., McIntyre V., Durouchoux P., Sood R., 2000, A\&A, 359, 251
\bibitem[\protect\citeauthoryear{Corbel}{2002}]{Corbel02} Corbel S., Fender R.P., 2002, ApJ, 573, L35
\bibitem[\protect\citeauthoryear{Corbel}{2003}]{Corbel03} Corbel S., Nowak M.A., Fender R.P., Tzioumis A.K., Markoff S., 2003, A\&A, 400, 1007
\bibitem[\protect\citeauthoryear{Corbel}{2009}]{Coriat09} Coriat M., Corbel S., Buxton M. M., Bailyn C. D.,
Tomsick J. A., K\"ording E., Kalemci E., 2009, MNRAS, 400, 123
\bibitem[\protect\citeauthoryear{Drury}{1983}]{Drury83} Drury L., 1983, Rep. Prog. Phys., 46, 973 
\bibitem[\protect\citeauthoryear{Dunn}{2008}]{Dunn} Dunn R.J.H., Fender R.P., K\"ording E.G., Cabanac C., Belloni T., 2008, MNRAS, 387, 545
\bibitem[\protect\citeauthoryear{FB}{1995}]{FB95} Falcke H., Biermann P. L., 1995, A\&A, 293, 665
\bibitem[\protect\citeauthoryear{Fender}{2004}]{Fender04} Fender R.P., Belloni T.M., Gallo E., 2004, MNRAS, 355, 1105 
\bibitem[\protect\citeauthoryear{FM}{2009}]{FM09} Fragile P.C., Meier D.L., 2009, ApJ, 693, 771
\bibitem[\protect\citeauthoryear{Gaisser}{1990}]{Gaisser90} Gaisser T. K., 1990, Cosmic Rays and Particle Physics. Cambridge University Press, Cambridge
\bibitem[\protect\citeauthoryear{Gallo}{2004}]{Gallo04}Gallo E., Corbel S., Fender R.P., Maccarone T.J., Tzioumis A.K., 2004, MNRAS, 347, L52
\bibitem[\protect\citeauthoryear{GS}{1967}]{GS67} Gould R.J., Schr\'eder G.P., 1967, Phys. Rev., 155, 1404 
\bibitem[\protect\citeauthoryear{Guessoum}{2006}]{G06} Guessoum N., Jean P., Prantzos N., 2006, A\&A, 2006, 457, 753
\bibitem[\protect\citeauthoryear{Hannikainen}{1998}]{Hannikainen05} Hannikainen D.C., Hunstead R.W., Campbell-Wilson D., Sood R.K., 1998, A\&A, 337, 460
\bibitem[\protect\citeauthoryear{Heinz}{2008}]{Heinz08} Heinz S., 2008, Int. J. Mod. Phys. D, 17, 1947
\bibitem[\protect\citeauthoryear{Heinz}{2003}]{Heinz03} Heinz S., Sunyaev R.A,  2003, MNRAS, 343, L59
\bibitem[\protect\citeauthoryear{Homan}{2005}]{Homan05} Homan J., Buxton M., Markoff S., Bailin C.D., Nespoli E., Belloni T., 2005, ApJ, 624, 295 
\bibitem[\protect\citeauthoryear{Hynes}{2003}]{Hynes03} Hynes R.I., Steeghs D., Casares J., Charles P.A., O'Brien K., 2003, ApJL, 583, L95
\bibitem[\protect\citeauthoryear{Hynes}{2004}]{Hynes04} Hynes R.I., Steeghs D., Casares J., Charles P.A., O'Brien K., 2004, ApJ, 609, 317
\bibitem[\protect\citeauthoryear{Imamura}{1990}]{Imamura90} Imamura J.N., Kristian J., Middleditch J., Steiman-Cameron T.Y., 1990, ApJ, 365, 312
\bibitem[\protect\citeauthoryear{Kay}{1993}]{Kay93} Kay S.M., 1993, Fundamentals of Statistical Signal Processing, Volume 1: Estimation Theory. Prentice Hall Signal Processing Series, Upper Saddle River, New Jersey 
\bibitem[\protect\citeauthoryear{Kelner}{2006}]{Kelner06} Kelner S.R., Aharonian F.A., Bugayov V.V., 2006, Phys. Rev. D, 74, 034018
\bibitem[\protect\citeauthoryear{Kelner}{2008}]{Kelner08} Kelner S.R., Aharonian F.A., 2008, Phys. Rev. D, 78, 034013
\bibitem[\protect\citeauthoryear{Keshet}{2005}]{Keshet05} Keshet U., Waxman E., 2005, Phys. Rev. Lett., 94, 111102  
\bibitem[\protect\citeauthoryear{Khangulyan}{2008}]{Khangulyan08}Khangulyan D., Aharonian F.A., Bosch-Ramon V., 2008, MNRAS, 383, 467
\bibitem[\protect\citeauthoryear{Krolik}{1999}]{Khangulyan07} Khangulyan D., Hnatic S., Aharonian F.A., Bogovalov S., 2007, MNRAS, 380, 320
\bibitem[\protect\citeauthoryear{Komissarov}{2007}]{Komissarov07} Komissarov S., Barkov M., Vlahakis N., K\"onigl A., 2007, MNRAS, 380, 51
\bibitem[\protect\citeauthoryear{Krolik}{2007}]{Krolik99} Krolik J.H., 1999, Active Galactic Nuclei. Princeton University Press, Princeton 
\bibitem[\protect\citeauthoryear{Levine}{2006}]{Levine06} Levine A.M., Corbet R., 2006, ATel, 940
\bibitem[\protect\citeauthoryear{Lipari}{2007}]{Lipari07} Lipari P., Lusignoli M., Meloni D, 2007, Phys. Rev. D, 75, 123005
\bibitem[\protect\citeauthoryear{Mannheim}{1994}]{Mannheim94} Mannheim K., Schlickeiser R., 1994, A\&A, 286, 983
\bibitem[\protect\citeauthoryear{Markert}{1973}]{Markert73} Markert T.H., Canizares C.R., Clark G.W., 1973, ApJ, 184, L67
\bibitem[\protect\citeauthoryear{Markoff}{2003}]{Markoff03} Markoff S., Nowak M.A., Corbel S., Fender R., Falcke H., 2003, A\&A, 397, 645
\bibitem[\protect\citeauthoryear{Markoff}{2005}]{Markoff05} Markoff S., Nowak M.A., Wilms J., 2005, ApJ, 635, 1203
\bibitem[\protect\citeauthoryear{Maximon}{1968}]{Maximon68} Maximon L.C., 1968, J. Res. Natl. Bur. Stand., 72B, 79
\bibitem[\protect\citeauthoryear{McClintock}{2006}]{McClintock06} McClintock J.E., Remillard R.A. 2006, in: Compact stellar X-ray sources, Walter Lewin \& Michiel van der Klis (Eds). Cambridge Astrophysics Series, No. 39. Cambridge University Press, Cambridge, p. 157
\bibitem[\protect\citeauthoryear{Mucke}{2000}]{Mucke00} M\"ucke A., Engel R., Rachen J.P., Protheroe R.J., Stanev T., 2000, Comm. Phys. Comp., 124, 290 
\bibitem[\protect\citeauthoryear{MunozD}{2008}]{MunozD08}Mu\~noz-Darias T., Casares J., Mart\'inez-Pais I.G., 2008, MNRAS, 385, 2205
\bibitem[\protect\citeauthoryear{NY}{1995}]{NY95} Narayan R., Yi I., 1995, ApJ, 444, 231
\bibitem[\protect\citeauthoryear{Nowak}{2002}]{Nowak02} Nowak M.A., Wilms J., Dove J.B., 2002, MNRAS, 332, 856
\bibitem[\protect\citeauthoryear{Nowak}{2005}]{Nowak05} Nowak M.A., Wilms J., Heinz S., Pooley G., Pottschmidt K., Corbel S., 2005, ApJ, 626, 1006
\bibitem[\protect\citeauthoryear{Orellana}{2007}]{Orellana07} Orellana M., Bordas P., Bosch-Ramon V., Romero, G. E., Paredes J. M., 2007, A\&A, 476, 9 
\bibitem[\protect\citeauthoryear{Reynoso}{2009}]{Reynoso09} Reynoso M.M., Romero G.E., 2009, A\&A, 493, 1 
\bibitem[\protect\citeauthoryear{Romero}{2008}]{Romero08} Romero G.E., Vila G.S, 2008, A\&A, 485, 623 
\bibitem[\protect\citeauthoryear{Romero}{2009}]{Romero09} Romero G.E., Vila G.S. 2009, A\&A, 494, L33
\bibitem[\protect\citeauthoryear{Russel}{2006}]{Russel06} Russell D. M., Fender R. P., Hynes R. I., Brocksopp C., Homan J.
Jonker P. G., Buxton M. M., 2006, MNRAS, 371, 1334
\bibitem[\protect\citeauthoryear{SS}{1973}]{SS73} Shakura N.I., Sunyaev R., 1973, A\&A, 24, 337
\bibitem[\protect\citeauthoryear{Schlickeiser}{2002}]{Schlickeiser02} Schlickeiser R., 2002, Cosmic Ray Astrophysics. Astronomy and Astrophysics Library Series, Springer, Berlin
\bibitem[\protect\citeauthoryear{Stecker}{2007}]{Stecker07} Stecker F.W., Baring M.G. Summerling E.J., 2007, ApJ, 667, L29-L32 
\bibitem[\protect\citeauthoryear{Weidenspointner}{2008}]{Weidenspointner08} Weidenspointner G. et al., 2008, Nature, 451, 159  
\bibitem[\protect\citeauthoryear{Wilms}{1999}]{Wilms09} Wilms J., Nowak M.A., Dove J.B., Fender R.P., Di Matteo T., 1999, ApJ, 522, 460
\bibitem[\protect\citeauthoryear{Yu}{2007}]{Yu07} Yu W., Lamb F.K., Fender R., Van der Klis M., 2007, ApJ, 663, 1309 
\bibitem[\protect\citeauthoryear{Zdziarski}{1998}]{Zdziarski98} Zdziarski A.A., Poutanen J., Mikolajewska J., Gierli\'nski M., Ebisawa K., Johnson W.N., 1998, MNRAS, 301, 435
\bibitem[\protect\citeauthoryear{Zdziarski}{2004}]{Zdziarski04} Zdziarski A.A., Gierli\'nski M., Mikolajewska J., Wardzi\'nski G., Smith D., Alan Harmon B., Kitamoto S., 2004, MNRAS, 351, 791

\end{thebibliography}
\end{document}